\documentclass[submission, Phys]{SciPost}
\usepackage{amssymb}
\usepackage{algorithm}
\usepackage{algpseudocode}

\hypersetup{
	colorlinks,
	linkcolor={red!50!black},
	citecolor={blue!50!black},
	urlcolor={blue!80!black}
}

\DeclareSymbolFont{usualmathcal}{OMS}{cmsy}{m}{n}
\DeclareSymbolFontAlphabet{\mathcal}{usualmathcal}

\usepackage{graphicx}
\usepackage{tikz}
\usepackage{amsmath}

\makeatletter
\newsavebox{\@brx}
\newcommand{\llangle}[1][]{\savebox{\@brx}{\(\m@th{#1\langle}\)}%
	\mathopen{\copy\@brx\kern-0.5\wd\@brx\usebox{\@brx}}}
\newcommand{\rrangle}[1][]{\savebox{\@brx}{\(\m@th{#1\rangle}\)}%
	\mathclose{\copy\@brx\kern-0.5\wd\@brx\usebox{\@brx}}}
\makeatother

\newcommand{\floor}[1]{\left\lfloor #1 \right\rfloor}
\newcommand{\tr}{\mathrm{Tr}}
\newcommand{\bra}[1]{\langle\,#1\,|}
\newcommand{\ket}[1]{|#1\rangle}

\newcommand{\Tr}{\mathrm{Tr}}

\definecolor{Ured}{HTML}{cc0000}
\definecolor{Ublue}{HTML}{1f65cf}
\definecolor{Ugreen}{HTML}{198a11}

\graphicspath{{figs/}}
\begin{document}

\begin{center}{\Large \textbf{
			Optimal compression of quantum many-body time evolution operators into brickwall circuits \\
}}\end{center}

\begin{center}
	Maurits S. J. Tepaske\textsuperscript{1,2},
	Dominik Hahn\textsuperscript{2} and
	David J. Luitz\textsuperscript{1,2$\star$}
\end{center}

\begin{center}
	\textsuperscript{1}Physikalisches Institut, Universit\"at Bonn, Nussallee 12, 53115 Bonn, Germany
	\\
	\textsuperscript{2}Max-Planck-Institut for the Physics of Complex Systems, \\N{\"o}thnitzer Straße 38, 01187
	Dresden, Germany
	\\
    ${}^\star$ {\small \sf \href{mailto:david.luitz@uni-bonn.de}{david.luitz@uni-bonn.de}}
\end{center}

\begin{center}
	\today
\end{center}

\section*{Abstract}
{\bf
    Near term quantum computers suffer from a degree of decoherence which is prohibitive 
    for high fidelity simulations with deep circuits. An economical use of circuit depth is 
    therefore paramount. For digital quantum simulation of quantum many-body systems, real 
    time evolution is typically achieved by a Trotter decomposition of the time evolution 
    operator into circuits consisting only of two qubit gates. To match the geometry of the 
    physical system and the CNOT connectivity of the quantum processor, additional SWAP gates 
    are needed. We show that optimal fidelity, beyond what is achievable by simple Trotter 
    decompositions for a fixed gate count, can be obtained by compiling the evolution operator 
    into optimal brickwall circuits for the $S=1/2$ quantum Heisenberg model on chains and ladders, 
    when mapped to one dimensional quantum processors without the need of additional SWAP gates. 
}

\vspace{10pt}
\noindent\rule{\textwidth}{1pt}
\tableofcontents\thispagestyle{fancy}
\enlargethispage{5mm}
\vspace{10pt}

\section{Introduction}\label{sec_intro}

Quantum processors are a rapidly evolving technology which is expected to be pivotal for 
many classically hard problems like integer factorization, database search, optimization 
and many others~\cite{Kitaev1995quantum,Grover1996Afast,Shor1997Polynomial,Ebadi2022Quantum}. 
While truly universal quantum computing is still a long shot, one of the most promising 
near-term applications is the simulation of complex quantum systems due to their relative 
similarity to the quantum hardware itself. The simulation of such systems on classical 
computers is extremely hard due to the exponential complexity in terms of storage and 
computer time, while both problems are naturally solved on quantum hardware.

There are two different approaches: analog and digital quantum simulators. Analog simulators 
are specifically engineered systems to mimic the corresponding dynamics of the target system 
and are often based on quantum optical setups. This technique has been successfully applied 
to condensed matter systems~\cite{feld_observation_2011,cocchi_equation_2016,bernien2017probing,Gross2017Quantum,Ebadi2022Quantum,scholl2021quantum} 
and lattice gauge theories~\cite{banuls2020simulating,mil2020scalable,yang2020observation} and 
is in principle extremely powerful but requires a tailored experimental setup for a given type 
of problem.

In contrast, digital quantum simulators~\cite{lloyd1996universal} rely on a discrete 
representation of the wave function on an array of two level systems (dubbed qubits), 
which can be fully controlled by a universal set of quantum gates which allows in 
principle for the representation of any unitary operation on the many-body wave function, 
represented as a sequence of gates. Due to the universal representation of the wave function, 
this is an attractive approach which is extremely flexible once a suitable mapping of the 
system of interest to qubits is devised. Recent applications include condensed matter 
systems~\cite{neill2016ergodic,salathe2015digital,las2014digital,lanyon2011universal,
richter_simulating_2021, variational_mansuroglu_2021}, 
simulations from quantum chemistry~\cite{aspuru2005simulated,mil2020scalable,kandala2017hardware,google2020hartree} 
and high-energy physics~\cite{martinez2016real, Funcke_2021}. Digital quantum simulations were 
also used to realize exotic phases of matter like time crystals~\cite{Frey2022realization,mi2022time} 
and quantum spin liquids~\cite{satzinger2021realizing}.

The state-of-the-art method for simulating the real time dynamics of complex quantum 
systems involves a factorization of the time evolution operator into a sequence of 
gates using Trotter decompositions of different orders~\cite{trotter1959product,suzuki1976generalized,suzuki1985decomposition,suzuki1990fractal,suzuki1991general}, 
introducing discrete time steps to get an approximation of the exact time evolution of the system. 
This introduces a discretization error, which can be systematically controlled by using 
smaller step sizes. As a downside, small step sizes require a larger number of gates. 
Due to the fragility of the quantum state stored in the machine, and due to hardware 
imperfections, each additional gate potentially introduces new sources of error due 
to dissipation processes. Hence a trade-off between discretization errors and errors 
due to intrinsic machine noise during the simulation is required. To achieve optimal 
fidelity in light of this tradeoff, it is therefore important to minimize the resource 
costs for a given simulation. Recent work yielded tighter bounds for the discretization 
errors~\cite{Childs2021Theory}. Furthermore, it was also argued recently that beyond a 
certain step size the fidelity of the Trotter decomposition breaks down in a universal 
fashion, leading to a regime of quantum chaos~\cite{Heyl2019Quantum,Kargi2021Quantum}. 
This sets also upper bounds for possible step sizes. It remains however unclear, whether 
better alternatives to Trotter decompositions exist.

One promising approach in this regard are quantum variational algorithms. The main idea 
of them is to approximate a time-evolved state using a parametrized 
circuit~\cite{benedetti2021hardware,Bolens2021Reinforcement,barison2021efficient,
berthusen2021quantum}. 
The parameters are then fixed using optimization algorithms on a quantum computer. Recent 
numerics suggest that the number of parameters needed to describe time-evolved states or 
ground states scales favorable even in comparison to 
matrix-product states~\cite{lin_real_2021,haghshenas2022variational}. Most of these algorithms 
involve optimization where gradients are measured directly on the quantum devices, or they 
use deep learning approaches. However, the measurement of gradients on a quantum device 
is currently infeasible due to the high error rates, while optimization using deep neural 
networks is not controlled.

In this paper we take a more universal approach. Rather than focussing on the wave function, 
we directly target the time evolution operator, aiming at a compact representation as a shallow 
circuit. We use brickwall circuits in which the gates are parametrized two qubit unitaries, 
connecting neighboring qubits in the architecture of the quantum processor as an ansatz for 
the time evolution operator. This parametrized circuit can be optimized classically to represent 
the time evolution operator for a given time step with high fidelity. The resulting circuit can 
then be repeated to evolve the quantum state to later times. We show that such an optimized 
circuit can yield significantly higher fidelity time evolution for a fixed gate count 
compared to the traditional Trotter decomposition and is thus superior for digital quantum 
simulation. 

We also show that this strategy allows us to obtain similar accuracy using significantly 
less gates, even for systems where the physical geometry does not coincide with the proposed 
circuit architecture, essentially ``baking in'' the otherwise required SWAP gates to match 
geometries into the circuit. As an interesting benchmark problem, we use our approach to 
compute out-of-time-ordered correlators (OTOCs) and show that we achieve better accuracy 
than Trotter methods with similar resource cost. Finally, we analyze the gate structure of 
the optimized gates, as a first step towards further improvements of this approach.

\section{Model and Method}\label{sec_method}

\subsection{Model}

\begin{figure}
	\centering
    \begin{minipage}{0.49\textwidth}
        \centering
        \includegraphics[width=0.99\columnwidth]{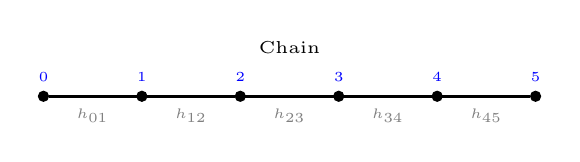}
\end{minipage}
    \begin{minipage}{0.49\textwidth}
        \centering
	\includegraphics[width=0.99\columnwidth]{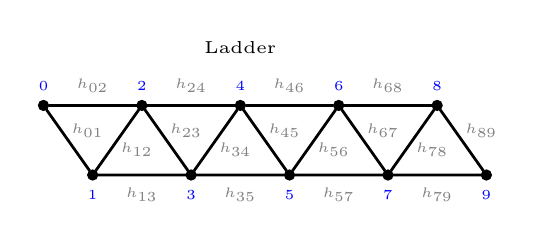}
\end{minipage}
    \caption{The chain (left) and triangular ladder (right) lattice geometries used in this work.}\label{lattices}
\end{figure}

For concreteness and simplicity, we focus on simulating finite systems of $s=1/2$ spins 
on a lattice with $L$ sites, designed to be performed on a quantum processor with an 
identical Hilbert space $\mathcal{H}$, which is the product space of $L$ two-level 
quantum systems (qubits) $\bigotimes_{i=1}^L\mathcal{Q}_i$ and has an exponentially 
growing dimension $\dim{H}=2^L$. Specifically, we discuss spin-$1/2$ systems with 
SU($2$) symmetric Heisenberg couplings
\begin{equation}
h_{ij} = \sigma^x_i\sigma^x_j + \sigma^y_i\sigma^y_j + \sigma^z_i\sigma^z_j,
\label{bond_hamiltonian}
\end{equation}
between nearest neighbor (NN) spins on a chain $\mathfrak{c}$ and a triangular ladder 
$\mathfrak{l}$, both with open boundary conditions, i.e.
\begin{equation}
H_\mathfrak{c} = \sum_{\langle i,j\rangle} h_{ij} \quad\quad\quad H_\mathfrak{l}=\sum_{\llangle i,j\rrangle} h_{ij}.
\label{hamiltonians}
\end{equation}
Here $\sigma_{x,y,z}$ are the usual Pauli matrices while $\langle i,j\rangle$ and 
$\llangle i,j\rrangle$ denote the NN sites of the chain, or the NN sites of our 
triangular ladder geometry (note that this is identical to a chain with nearest and 
next nearest neighbor (NNN) interactions). These lattice geometries are illustrated 
in Fig. \ref{lattices}. 

\begin{figure}
	\centering
    \begin{minipage}{0.59\textwidth}
        \centering
        \includegraphics[width=0.9\columnwidth]{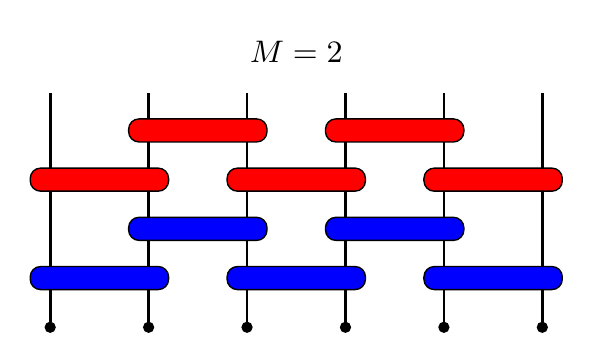}
\end{minipage}
    \begin{minipage}{0.39\textwidth}
        \centering
	\includegraphics[width=0.5\columnwidth]{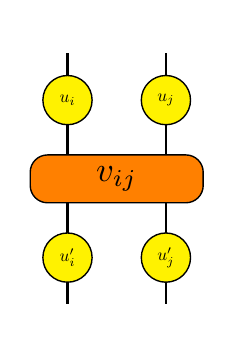}
\end{minipage}
\caption{Left: A brickwall circuit with depth $M=2$ for six qubits, with each color representing 
	             a $M=1$ layer. Circles represent the initial state of the qubits and boxes indicate 
	             a two qubit unitary gate applied to a pair of neighboring qubits. 
	       Right: Parametrization of a two qubit unitary as a product of four single qubit gates 
	              and one two qubit gate.}\label{bw_circuit}
\end{figure}

Most current quantum devices using superconducting qubits are not capable of all-to-all 
connectivity, i.e. due to the chip setup two qubit gates can only be applied between neighboring 
qubits, which are arranged in different geometries~\cite{arute2019quantum,IBM,Rigetti} 
In order to apply gates between distant qubits, one has to use a sequence of swap gates, which 
exchange the quantum state of neighboring qubits, such that effectively the states of distant 
qubits are moved to neighboring qubits in the processor geometry. On these, any two qubit gate 
can be applied and then the swap sequence needs to be applied in reverse order.
This requires a great number of additional gates and therefore introduces further possible sources 
of errors. 
 
Our goal is therefore to find the best unitary circuit $\mathcal{C}$ of a given depth $M$ to 
approximate the time evolution operator $\mathcal{U}(t)=\exp(-itH_{\mathfrak{c}/\mathfrak{l}})$. 
In order to mimic the limited connectivity of current quantum devices, we choose $\mathcal{C}$ 
to consist only of NN two-qubit gates on a 1d chain, arranged in a brickwall pattern, i.e. 
we model our quantum processor as an open chain of qubits, while one of our physical models 
we want to simulate on this machine has a different, triangular ladder, geometry. This allows 
us to investigate whether it is possible to compile the time evolution operator in a nearest 
neighbor, brickwall circuit (exemplified in the left panel of Fig. \ref{bw_circuit}) without the 
need for additional swap gates, which are generally costly on superconducting platforms.

\subsection{Trotter circuits}\label{sec:trotter}

\begin{figure}
	\centering
    \begin{minipage}{0.49\textwidth}
        \centering
        \includegraphics[width=0.9\columnwidth]{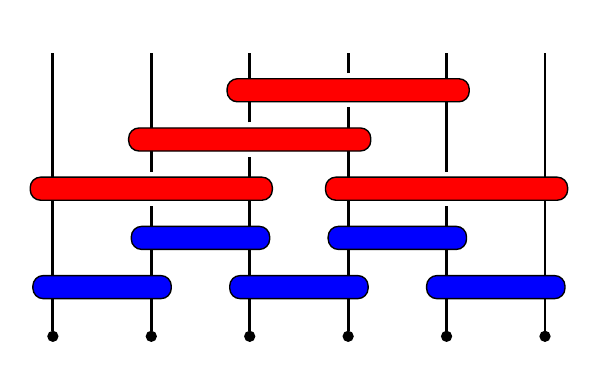}
\end{minipage}
    \begin{minipage}{0.49\textwidth}
        \centering
	\includegraphics[width=0.99\columnwidth]{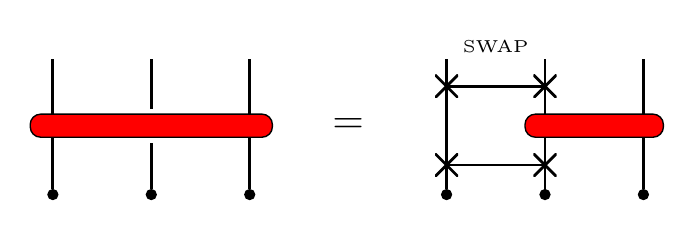}
\end{minipage}
\caption{Left: The blue brickwall layer encodes the first-order Trotter decomposition for NN 
	             interachting Hamiltonians. The combination of the blue and red layers encodes 
	             the first-order Trotter decomposition for NNN interacting Hamiltonians, where the 
	             blue gates act on NN qubits whereas the red gates act on NNN qubits. 
	       Right: The decomposition involving SWAP gates, displayed as the crossed line, which is 
	              used to convert the NNN two-qubit gate into a circuit involving only two-qubit 
	              gates.}\label{nnn_trotter}
\end{figure}

To benchmark the performance of the brickwall circuits we will compare them with the first-, 
second- and fourth-order Trotter circuits that are based on the well known Trotter 
decompositions \cite{paeckel_time_2019}. Here we introduce these circuits for the 
Hamiltonians (\ref{hamiltonians}) that are used in this work.

For the chain Hamiltonian $H_\mathfrak{c}$ we have two non-commuting parts, namely the bond 
Hamiltonians $h_{i,i+1}$ (\ref{bond_hamiltonian}) on alternating bonds, such that we can 
split $H_\mathfrak{c}$ in two commuting parts as 
\begin{equation}
H_\mathfrak{c}=H_1+H_2=\sum_{i=0,2,\dots}h_{i,i+1} + \sum_{i=1,2,\dots}h_{i,i+1}.
\end{equation}
For the ladder Hamiltonian we have on top of this three extra non-commuting parts due to the 
NNN couplings, i.e. we can split $H_\mathfrak{l}$ into five commuting parts as 
\begin{equation}
H_\mathfrak{l}=H_\mathfrak{c}+H_3+H_4+H_5=H_\mathfrak{c}+\sum_{i=0,3,\dots}h_{i,i+2} + \sum_{i=1,4,\dots}h_{i,i+2} + \sum_{i=2,5,\dots}h_{i,i+2}.
\end{equation}
By writing the Hamiltonians in this way we can define the $M=1$ first-order Trotter circuits 
for $H_\mathfrak{c}$ and $H_\mathfrak{l}$ as \cite{paeckel_time_2019}
\begin{align}
\mathcal{U}^\mathrm{1st}_\mathfrak{c}(t)&=\mathcal{U}_1(t)\mathcal{U}_2(t), \\
                                      &= \exp(-\mathrm{i}tH_1)\exp(-\mathrm{i}tH_2), \\
\mathcal{U}^\mathrm{1st}_\mathfrak{l}(t)&=\mathcal{U}_1(t)\mathcal{U}_2(t)\mathcal{U}_3(t)\mathcal{U}_4(t)\mathcal{U}_5(t), \\
                                      &= \exp(-\mathrm{i}tH_1)\exp(-\mathrm{i}tH_2)\exp(-\mathrm{i}tH_3)\exp(-\mathrm{i}tH_4)\exp(-\mathrm{i}tH_5).
\end{align}
These circuits approximate the exact $\mathcal{U}(t)=\exp(-\mathrm{i}t H_{\mathfrak{c}/\mathfrak{l}})$ 
with error $\mathcal{O}(t^2)$~\cite{Childs2021Theory}. Note that depth $M=1$ for the Trotter 
circuits does not mean one brickwall layer, but instead one Trotter step 
$\mathcal{U}^\mathrm{1nd}_{\mathfrak{c}/\mathfrak{l}}(t)$. While these coincide for the first-order 
Trotter circuit for the chain, this is not the case for the first-order Trotter circuit for the 
ladder, and for the second- and fourth-order Trotter circuits which we introduce below. The 
circuit diagram for $\mathcal{U}^\mathrm{1st}_\mathfrak{c}(t)$ is shown as the blue brickwall 
layer in the left panel of Fig. \ref{nnn_trotter}, where $\mathcal{U}_1(t)$ is the half-brickwall 
layer on odd bonds and $\mathcal{U}_2(t)$ is the half-brickwall layer on even bonds. The circuit 
diagram for $\mathcal{U}^\mathrm{1st}_\mathfrak{l}(t)$ is the full circuit in this figure, 
where $\mathcal{U}_1(t)$ and $\mathcal{U}_2$ again form the blue brickwall layer while 
$\mathcal{U}_3(t)$, $\mathcal{U}_4(t)$ and $\mathcal{U}_5(t)$ form the red layer, containing 
two-qubit gates that act on NNN instead of NN qubits. To turn this into a circuit that involves 
only NN two-qubit gates we introduce the SWAP gate and decompose every NNN gate as in the right 
panel of Fig. \ref{nnn_trotter}.

The circuit layers $\mathcal{U}_1,\mathcal{U}_2,\mathcal{U}_3,\mathcal{U}_4,\mathcal{U}_5$ form 
the building blocks of the second- and fourth-order Trotter circuits. The $M=1$ second-order 
Trotter circuits are composed as \cite{paeckel_time_2019}
\begin{align}
\mathcal{U}^\mathrm{2nd}_\mathfrak{c}(t)&=\mathcal{U}_1(t/2)\mathcal{U}_2(t)\mathcal{U}_1(t/2), \\
\mathcal{U}^\mathrm{2nd}_\mathfrak{l}(t)&=\mathcal{U}_1(t/2)\mathcal{U}_2(t/2)\mathcal{U}_3(t/2)\mathcal{U}_4(t/2)\mathcal{U}_5(t)\mathcal{U}_4(t/2)\mathcal{U}_3(t/2)\mathcal{U}_2(t/2)\mathcal{U}_1(t/2),
\end{align}
which approximate the exact evolution operators with error 
$\mathcal{O}(t^3)$~\cite{Childs2021Theory}. Using these second-order Trotter circuits we can 
define the $M=1$ fourth-order Trotter circuits as \cite{paeckel_time_2019}
\begin{equation}
\mathcal{U}^\mathrm{4th}_{\mathfrak{c}/\mathfrak{l}}(t)=\mathcal{U}^\mathrm{2nd}_{\mathfrak{c}/\mathfrak{l}}(t_1)\mathcal{U}^\mathrm{2nd}_{\mathfrak{c}/\mathfrak{l}}(t_1)\mathcal{U}^\mathrm{2nd}_{\mathfrak{c}/\mathfrak{l}}(t_2)\mathcal{U}^\mathrm{2nd}_{\mathfrak{c}/\mathfrak{l}}(t_1)\mathcal{U}^\mathrm{2nd}_{\mathfrak{c}/\mathfrak{l}}(t_1),
\end{equation}
where we defined the time steps
\begin{equation}
t_1=\frac{1}{4-4^{1/3}}t \quad\quad\quad t_2 = (1-4t_1)t.
\end{equation}
These circuits approximate the exact evolution operators with error 
$\mathcal{O}(t^5)$~\cite{Childs2021Theory}.

Because we are concerned with circuits that are implemented on a quantum processor with only 
NN qubit connectivity, we have to convert every NNN two-qubit gate that appears in 
$\mathcal{U}^\mathrm{1st}_\mathfrak{l},\mathcal{U}^\mathrm{2nd}_\mathfrak{l},\mathcal{U}^\mathrm{4th}_\mathfrak{l}$ 
to three NN two-qubit gates, as shown in the right panel of Fig. \ref{nnn_trotter}. The gate 
counts $N_g$ of the resulting NN Trotter circuits are given in Sec. \ref{sec_gate_counts}, also 
for the chain geometry.

\subsection{Optimization}\label{sec:optimization}

Each two-qubit gate $U_{ij}\in\mathbb{C}^{4\times 4}$ of the circuit $\mathcal C$, acting on two 
neighboring qubits $i$ and $j$, can be decomposed into a product of one-qubit gates 
$u_i \in \mathbb{C}^{2\times 2}$ and a two-qubit gate $v_{ij}\in \mathbb{C}^{4\times 4}$ \cite{kraus_optimal_2001} 
\begin{equation}
U_{ij} = (u_i\otimes u_j) v_{ij} (u'_i\otimes u'_j).
\label{gate_param}
\end{equation}
Here $v_{ij}$ is parameterized as
\begin{equation}
v_{ij}(\lambda_0,\lambda_1,\lambda_2) = e^{-i(\lambda_0\sigma^x_i\otimes\sigma^x_j + \lambda_1\sigma^y_i\otimes\sigma^y_j + \lambda_2\sigma^z_i\otimes\sigma^z_j)},
\label{twoqubit_param}
\end{equation}
with three real parameters $\lambda_{0,1,2}$, and the $u_i$ are parameterized up to a global 
phase as
\begin{equation}
u_i(\phi_0,\phi_1,\phi_2) = 
\begin{pmatrix}
e^{i\phi_1}\cos(\phi_0) & e^{i\phi_2}\sin(\phi_0) \\
-e^{-i\phi_2}\sin(\phi_0) & e^{-i\phi_1}\cos(\phi_0),
\end{pmatrix},
\label{onequbit_param}
\end{equation}
each containing three real parameters $\phi_{0,1,2}$. Hence this decomposition of $U_{ij}$ 
contains $15$ real parameters, and it can be visualised as in the right panel of 
Fig. \ref{bw_circuit}. To represent the unitary gate as a global unitary matrix, acting on the 
full wave function, we introduce its matrix form
\begin{equation}
    \text{mat}(U_{ij}) = I_{2^{i-1}} \times U_{ij} \times I_{2^{L-j}},
\end{equation}
by taking the Kronecker product with identity matrices on the qubits on which the gate does not 
act (and implicitly encoding the nearest neighbor condition $j=i+1$).
The entire circuit is a product of such unitaries and can formally be expressed by
\begin{equation}
    \mathcal{C} = \prod_{k=0}^{N_g} \text{mat}( U_{i_k, j_k} ),
    \label{eq:circ}
\end{equation}
where $N_g$ is the total number of gates in the circuit. Since each gate is parametrized by 
$\vec \theta_{i_k}= (\vec \lambda_{i_k}, \vec \phi_{i_k})$, the circuit depends on all these 
$15 N_g$ parameters $\vec \theta = (\vec \theta_{i_0}, \vec \theta_{i_1}\dots) \in \mathbb{R}^{15 N_g}$
\begin{equation}
    \mathcal{C}(\vec \theta) = \prod_{k=0}^{N_g} \text{mat}( U_{i_k, j_k}(\vec \theta_{i_k}) ).
    \label{eq:circ2}
\end{equation}
In practice, when stacking the gates to form the circuit, we merge two one-qubit unitaries into a 
single one-qubit unitary where possible, since a product of general one-qubit unitaries can be written
as a single general one-qubit unitary. This reduces the amount of circuit parameters.

We would like to find an optimal parameter set $\vec \theta$ for a given circuit architecture, 
such that the distance between the unitary represented by the circuit $\mathcal C(\vec \theta)$ 
and the targeted time evolution operator $\mathcal U$ of the system up to time $t$ is minimized. 
For two unitary operators $\mathcal{U}$ and $\mathcal{C}$, we therefore define a measure of 
distance in terms of the normalized Frobenius norm, namely the "infidelity" $\epsilon$, given by
\begin{equation}
\epsilon = \frac 1 2 \frac{\| \mathcal{U} - \mathcal{C} \|_F^2}{2^L} = \frac{1}{2^{L+1}} \tr\big[ (\mathcal{U} - \mathcal{C})^\dagger (\mathcal{U} - \mathcal{C}) \big] = 1 - \frac{ \mathrm{Re} \tr \big[\mathcal{U}^{\dagger}\mathcal{C}\big] }{2^L}.
    \label{eq:fidelity_def}
\end{equation}
We use this infidelity as an objective function, such that we obtain a minimization problem for a 
fixed circuit architecture (number and sequence of two qubit gates). In our case the target 
unitary $\mathcal U$ is an approximation of an exact time-evolution operator, where the error 
stems from the tensor network methods that make the optimization tractable.

The objective function $\epsilon$ needs to be evaluated many times during the optimization and 
we find that it is efficient to first compress the time evolution operator $\mathcal U$ into a 
matrix product operator (MPO) $\mathcal{M}_\chi$ of bond dimension $\chi$, such that we can 
calculate $\epsilon$ via efficient standard tensor network methods. For the local systems we 
investigate here and for short times, this is always efficient, due to the low operator 
entanglement of the time evolution operator \cite{zhou_operator_2017}. In particular, we discard 
the smallest singular values for which the squares sum to a tiny number, since their contribution 
is negligible, such that lowly entangled operators do not saturate the maximum bond dimension 
$\chi$. To obtain the (truncated) MPO representation of $\mathcal{U}$ with negligible 
discretization error, we take an identity MPO and perform time-evolving block 
decimation \cite{vidal_efficient_2003, paeckel_time_2019} with a small timestep 
$\delta t=10^{-4}$ and fourth-order Trotter decomposition, such that the introduced error is 
negligible\footnote{We compare the results for our circuits to Trotter circuits with comparable 
gate counts, and in all instances of the involved Trotter circuits, the timesteps are several 
orders of magnitude larger than the stepsize used to approximate the target 
unitary $\mathcal U$.}. 

To optimize the parameters $\vec{\theta}$ of the circuit such that $\epsilon$ is minimal, 
we employ the paradigm of differentiable programming \cite{liao_differentiable_2019}. 
Here the gradient $\nabla_{\vec{\theta}}\, \epsilon$ is calculated in a similar fashion as the 
original backpropagation algorithm used for deep neural networks \cite{rumelhart_learning_1986}, 
which has been generalized to arbitrary programs, including tensor network algorithms \cite{liao_differentiable_2019}.
To this end, a program is represented as a computational graph through which the local gradients are propagated, 
which requires each computational component to have a well-defined gradient. In particular, for the tensor network algorithm 
in this work, the SVD is a crucial component, and so it is important to construct a stable SVD gradient \cite{liao_differentiable_2019}.
Fortunately, differentiable programming inherits the cost from its base algorithm, i.e. in our case
from the $M$ SVDs that are performed when obtaining the circuit MPO at every iteration. As a result 
our algorithm has the scaling $\mathcal{O}(N_i L M d^6 \chi^3)$, where $N$ is the amount of 
gradient descent iterations. Importantly, even though the cost scales linearly with system size $L$ 
and circuit depth $M$, the amount of parameters grows as $\mathcal{O}(L M)$, such that the amount of iterations
required to reach a low-lying minimum also grows, because local minima prolifrate with growing parameter count \cite{sack_transition_2022}. 

Using the global gradient $\nabla_{\vec{\theta}}\, \epsilon$ we then perform gradient descent. 
We use this global optimization procedure instead of the local optimization from \cite{lin_real_2021} 
because we found that this yields significantly higher fidelity when an Adam-like adaptive 
learning rate is used \cite{adam_kingma_2014}. Here it is crucial not to stop optimizing when 
the infidelity appears to have stagnated, since we have often found that the optimization 
gets stuck in such a "local minimum" for some time before it jumps out and converges to a 
lower minimum. This is possibly related to the "barren plateau" problem that often occurs when 
performing gradient descent for quantum circuits with a large parameter space, where the 
optimization reaches a set of circuit parameters for which the majority of its gradients become 
very small such that the optimization (temporarily) halts \cite{mcclean_barren_2018}. In 
Sec. \ref{sec:convergence} we review the Adam method and discuss the mentioned convergence 
behavior in more detail.

At small $M$ the optimized circuits in a sense compress the targeted time evolution operator, 
especially when its time-step is large, and therefore they are called "compressed circuits". In 
Sec. \ref{sec:lattice_symmetries} we check if the lattice symmetries of the targeted unitary 
emerge in the gates of the optimized circuits.

\subsection{Stacking circuits}

The general strategy we implement is the following: For some (short) timestep $t$, we find an 
optimal circuit $\mathcal C(\vec \theta)$ which best approximates the targeted time evolution 
operator $U_t$. In principle, $t$ is arbitrary, with the general logic that shorter $t$ unitaries 
can be encoded by shallower circuits (lower $M$). In practice, $t$ will be also governed by the 
time grid, on which observables should be evaluated, although this could be achieved also by 
working with two or more different optimized circuits with different $t$, a case we do not further 
discuss in this work. To propagate the wave function to longer times, which are multiples of $t$, 
we then use the circuit
\begin{equation}
    \mathcal C(\vec \theta)^{n} \approx \mathcal U_t^{n}.
\end{equation}
It is interesting to investigate how well this stacked circuit performs for time evolution to 
longer times and we will confront these results to benchmarks for the circuits discussed in 
Sec. \ref{sec:trotter} that result from traditional Trotter decompositions.

\subsection{Quantities of interest}

Having obtained the compressed circuits for short times, for which the 
relatively low entanglement allows for an accurate description with truncated MPOs,
we then compute $\epsilon$ for long times using the stacked circuits as approximation. 
If we now were to use the same MPO formalism that was used during the optimization, 
the growing of entanglement as we stack the circuit multiple times results in either 
an unfeasible amount of required computational resources or significant truncation errors. 
In particular, the stacked circuit represents a target unitary at large times, which generally 
has large entanglement, such that an accurate MPO representation requires a saturated bond 
dimension, i.e. the central tensors would require bond dimension $2^L$ to prevent significant 
truncation errors. 

For a highly entangled MPS $\ket{\psi_i}$ this central bond dimension is 
instead $2^{L/2}$, which is still managable for the system sizes considered in 
this work. Hence, to probe the true representablity of the stacked circuit, without having to deal with 
artefacts of the tensor network method, we use typicality~\cite{luitz_information_2017}. Here 
the trace in Eq. \ref{eq:fidelity_def} is replaced by the average over $N_\psi$ Haar random states 
$\ket{\psi_i}$, i.e.
\begin{equation}
\Tr\big[\mathcal{U}^{\dagger}\mathcal{C}\big] \approx \frac{1}{N_\psi}\sum_i \bra{\psi_i}\mathcal{U}^{\dagger}\mathcal{C}\ket{\psi_i}.
\label{eq_typicality}
\end{equation}
This allows us to calculate $\epsilon$ in an unbiased manner for the system sizes considered in 
this work. 

Besides using the infidelity $\epsilon$ as a measure of the performance of the circuits, we will 
also use the circuits to compute out-of-time-ordered correlators (OTOCs) \cite{hemery_matrix_2019}. 
For spin-$1/2$ $\sigma^z$ operators, the OTOC $C_{ij}$ between lattice sites $i$ and $j$ is 
defined with the Frobenius norm as
\begin{equation}
C_{ij}(t) = \left\Vert\left[ \sigma^z_i(t), \sigma^z_j \right]\right\Vert^2_\mathrm{{F}},
\label{otoc_definition}
\end{equation}
where $\sigma^z_i(t)=\mathcal{C}^{\dagger}\sigma^z_i\mathcal{C}$ is the spin operator on 
site $i$ evolved by the circuit. As for the infidelity, it is important to use typicality 
instead of the truncated MPO formalism when calculating $C_{ij}$ for a circuit that is stacked 
many times. 

To calculate (\ref{otoc_definition}) we invoke the hermiticity of the spin operators $\sigma^z$, 
such that by expanding the commutator in (\ref{otoc_definition}) we can write the OTOC as
\begin{equation}
C_{ij}(t) = 1-\frac{1}{4}\tr\big[\sigma_j^z\sigma^z_i(t)\sigma_j^z\sigma^z_i(t)\big],
\label{otoc_definition_reduced}
\end{equation}
which is readily calculated in the MPO formalism. Concretely, we take an identity MPO and put 
a $z$-spin operator $\sigma^z$ at site $i$, which is then evolved in the Heisenberg picture 
by the circuit $\mathcal{C}$, yielding a different MPO. Then we again take an identity MPO and 
put a $z$-spin operator on site $j$, which we do not evolve. Then we calculate the trace 
in (\ref{otoc_definition_reduced}) via a full contraction of four MPOs, which can be done 
efficiently.

\section{Results}\label{sec_results}

To benchmark the performance of the compression strategy outlined in Sec. \ref{sec_method}, we 
systematically analyze the infidelity $\epsilon$ as a function of simulation time step $t$, 
total gate count $N_g$ and system size, in direct comparison to Trotter decompositions of 
different orders, and present these results in Sec. \ref{subsec_infidelity}. In 
Sec. \ref{subsec_otocs} we extend this systematic analysis to out-of-time-ordered correlators 
(OTOCs) (\ref{otoc_definition}). Furthermore, in Sec. \ref{subsec_structure} we probe the 
structure of the gates that make up the optimized circuits, in an attempt to uncover the 
structures that allow these circuits to outperform their Trotter counterparts. 

\subsection{Infidelity}\label{subsec_infidelity}
\begin{figure*}[t!]
	\centering
	\includegraphics[width=\textwidth]{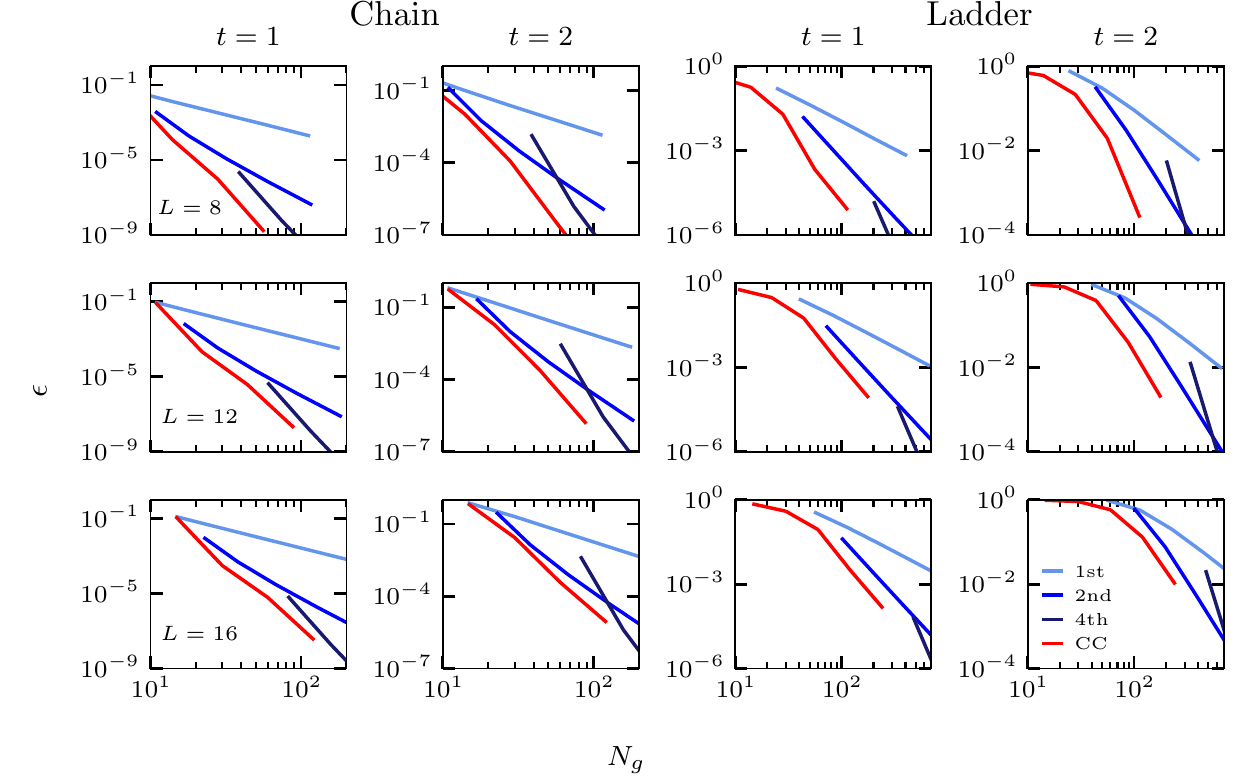}
	\caption{The infidelity $\epsilon$ versus gate count $N_g$ for the time evolution 
	         operator of the Heisenberg model on a chain (left panels) and ladder (right panels) 
	         in log-log scale. The first and third columns are for $t=1$ while the second and 
	         fourth columns are for $t=2$. The top panels are for $L=8$ and a time evolution MPO 
	         with $\chi=256$, the middle panels are for $L=12$ with $\chi=150$, and the bottom 
	         panels are for $L=16$ with $\chi=100$. The blue curves represent the Trotter circuits 
	         and the red curve represents the compressed circuit (CC).}
	\label{infidelity_Ng_dependency}
\end{figure*}
As a first test of the circuit optimization algorithm outlined in Sec.~\ref{sec_method}, we 
compare the optimal infidelities of compressed circuits to those of comparable Trotter circuits. 
Concretely, we consider time evolution operators of the chain and ladder Heisenberg Hamiltonians 
(\ref{hamiltonians}) at three system sizes $L=8,12,16$ and two time-steps $t=1,2$. For each 
Hamiltonian, system size and time-step, we determine the time evolution operator $\mathcal{U}$ 
with numerically negligible discretization error for a certain bond dimension $\chi$, and 
perform the global optimization as outlined in Sec \ref{sec_method} to minimize the infidelity 
$\epsilon$ of the compressed circuit. For $L=8,12,16$ we have taken $\chi=256,150,100$ as a 
compromise between precision and practical efficiency. We note that our main concern here is 
not to get a numerically exact MPO representation, but rather a reasonably good approximation of 
the time evolution operator. We call this our target time-evolution operator, which we want to 
approximate with our circuits.

As a first benchmark, we take for each of our parameter sets various circuit depths 
$M=1,2,4,8,16$, where $M$ is the number of elementary layers of $L-1$ gates, and consider 
$\epsilon$ as a function of the corresponding gate count $N_g$ (see Sec. \ref{sec_gate_counts} 
for details on how to obtain the number of gates). We compare this with first-, second- and 
fourth-order Trotter circuits \cite{paeckel_time_2019}.

The results are shown in Fig.~\ref{infidelity_Ng_dependency}. The left pair of panel columns 
is for the chain and the right pair is for the ladder. The first and third columns are for 
time-step $t=1$ and the second and fourth are for $t=2$. The upper row is for system size $L=8$, 
the middle row is for $L=12$, and the bottom row is for $L=16$. Each panel contains the 
infidelities of the optimized compressed circuits (CC) as a red line, and the infidelities of 
the Trotter circuits as blue lines. The infidelities of the Trotter circuits are calculated for 
the same depths $M$ as the compressed circuit, where it should be remembered from Sec. 
\ref{sec:trotter} that in this case $M$ is not necessarily equal to the amount of brickwall 
layers in the Trotter circuit, but is instead equal to the amount of Trotter steps that compose 
the circuit. The time-step of the Trotter step is chosen as $t/M$, such that $M$ subsequent steps 
correspond to a total time-step $t$. The gate counts of the Trotter circuits were calculated 
with the expressions in Sec.~\ref{sec_gate_counts}, which take into account the number of swap 
gates required to map the ladder geometry to a chain of qubits.

\begin{figure*}[t!]
	\centering
	\includegraphics[width=\textwidth]{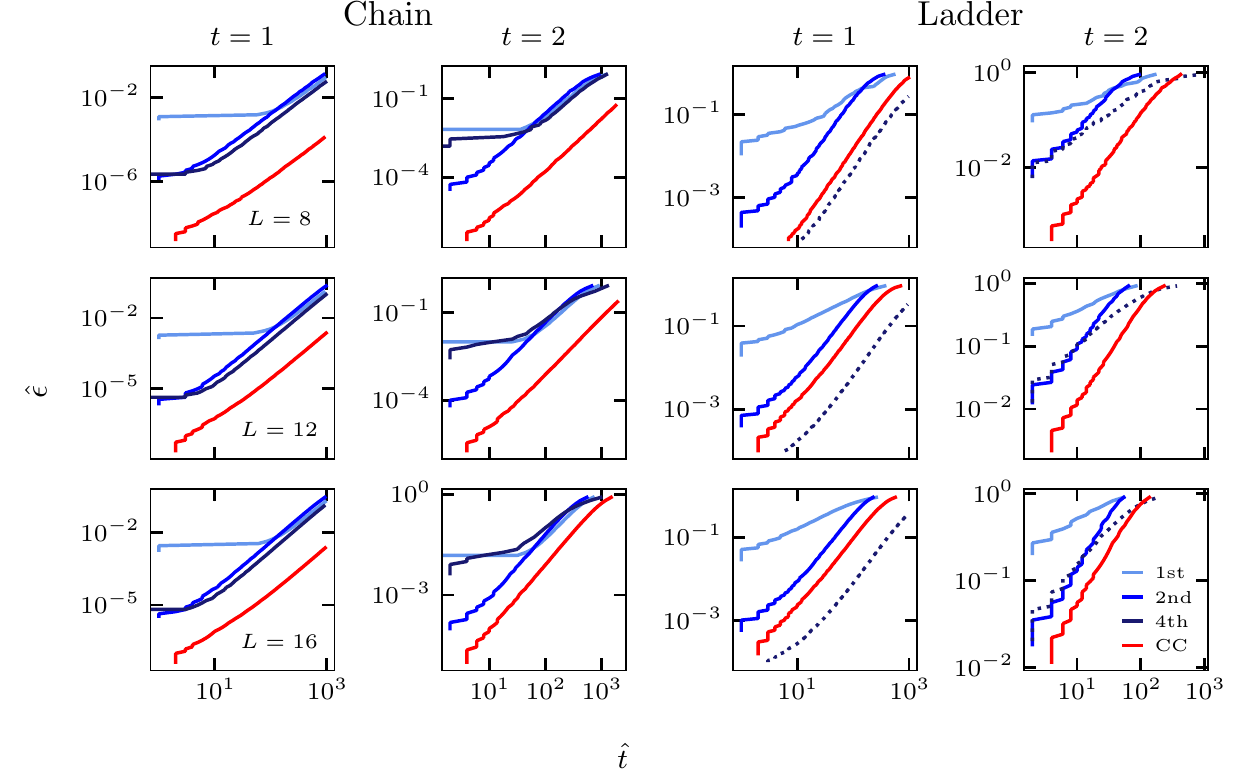}
	\caption{The time $\hat{t}$ after which the stacked circuits exceed the infidelity
	         threshold $\hat{\epsilon}$, for the time evolution operator of the Heisenberg 
	         model on a chain (left panels) and ladder (right panels) in log-log scale. 
	         The first and third columns are for circuits optimized at $t=1$ while the second 
	         and fourth columns are for $t=2$, with the circuits being stacked up to a thousand 
	         times. The circuits were chosen such that they have similar gate counts, with 
	         $M=8,8,7,1$ for the chain and $M=16,4,3,1$ for the ladder, for the compressed circuit 
	         and first-, second- and fourth-order Trotter circuits, respectively. The top panels 
	         are for $L=8$ with $\chi=256$, the middle panels are for $L=12$ with $\chi=150$, 
	         and the bottom panels are for $L=16$ with $\chi=100$. The blue curves represent the 
	         Trotter circuits and the red curve represents the compressed circuit (CC). The 
	         fourth-order Trotter circuit for the ladder is displayed as a dashed line, since 
	         it contains roughly twice as many gates as the compressed circuit and is therefore 
	         not necessarily indicative of their relative performance.}
	\label{infidelity_stack_threshold}
\end{figure*}

From Fig.~\ref{infidelity_Ng_dependency} it becomes clear that per gate the compressed circuit 
outperforms the Trotter circuits for all considered parameter sets. Moreover, it appears that 
for $L=8$ the infidelity of the compressed circuit roughly scales with $N_g$ like the best 
Trotter order, but with a more favorable prefactor, i.e. at intermediate gate counts it scales 
as second-order whereas at the highest probed gate count it scales as fourth-order. We have 
found that the same picture emerges when plotting $\epsilon$ versus the $t$ at which the circuit
was optimized, where $M=1$ scales like first-order Trotter, and by increasing $M$ we approach
the fourth-order scaling, passing through the second-order scaling. We expect the same to hold
for $L=12$ and $L=16$, if we could reach a lower minimum, but here the optimization is more
expensive.

Having considered the infidelities of the compressed circuits at the time-step for which they
were optimized, we now quantify how these infidelities grow when the circuits are stacked, 
which we do for the same systems as in Fig. \ref{infidelity_Ng_dependency}. To this end we 
select a compressed circuit that was optimized at $t=2$, and take for every Trotter order a 
circuit of depth $M$ with a gate count as close as possible to that of the compressed circuit, 
and choose its time-step to be $t/M$. 

Concretely, for the chain we take a compressed circuit with $M=8$, in which case we have to take 
first-, second-, and fourth-order Trotter circuits with $M=8,7,1$. Using the gate count equations 
from Sec. \ref{sec_gate_counts} we find that for $L=8$ the circuits have $N_g=56,56,53,39$, 
for $L=12$ they have $N_g=88,88,83,61$, and for $L=16$ they have $N_g=120,120,113,83$. For the 
ladder we take a compressed circuit with $M=16$, such that we have to take first-, second-, and 
fourth-order Trotter circuits with $M=4,3,1$. The corresponding gate counts are 
$N_g=112,100,124,204$ for $L=8$, $N_g=176,164,207,341$ for $L=12$, and $N_g=240,228,290,478$ for 
$L=16$.

To quantify the quality of the compressed and Trotter circuits under stacking, we take various 
infidelity thresholds $\hat{\epsilon}$ and stack the circuits up to a thousand times until they 
cross this threshold at some time $\hat{t}$, i.e. we determine $\epsilon(\hat{t})=\hat{\epsilon}$. 
As mentioned in Sec. \ref{sec_method} we utilize typicality (\ref{eq_typicality}) to calculate 
the stacked infidelities.

In Fig.~\ref{infidelity_stack_threshold} we plot $\hat{\epsilon}$ versus $\hat{t}$ in 
log-log scale. The used color coding is identical to that of Fig. \ref{infidelity_Ng_dependency}, 
except that the fourth-order Trotter circuit for the ladder is now represented with a 
dashed line, to emphasize that its infidelity relative to that of the compressed circuit 
is not necessarily indicative of the relative performance, because it contains roughly twice 
as many gates as the compressed circuit. From these plots it is clear that the advantage of 
the compressed circuits from Fig. \ref{infidelity_Ng_dependency} is not lost when stacking it 
many times. In particular, in all considered cases the compressed circuits are able to go to 
significantly larger times, at all infidelity thresholds, than the Trotter counterparts. The 
only exception is for the ladder at $t=1$, where the fourth-order Trotter circuit performs 
better, but as mentioned this Trotter circuit has twice as many gates as the compressed 
circuit and is therefore not a fair comparison. 

From the plots we extract the universal quadratic power-law $\hat{\epsilon}\propto\hat{t}^2$, 
for both the compressed and the Trotter circuits. This error scaling is analogous to first-order 
Trotter decomposition. The only exception is the ladder with $L=16$ at $t=2$, where the 
infidelity reaches $\epsilon\approx1$ rather quickly, such that it is situated in the rounding 
part that is also observed for the $t=1$ ladder curves at the high-infidelity end. The gap 
between the compressed circuits and the best performing Trotter circuits is thus found to grow 
quadratically with $\hat{t}$. Concretely, for the chain with $L=12$ and timestep $t=1$, we 
find that for $\hat{\epsilon}=10^{-3}$ the compressed circuit has $\hat{t}=644$ whereas the 
best Trotter circuit (i.e. of fourth-order) has $\hat{t}=94$. For $\hat{\epsilon}=10^{-4}$ 
we instead get $\hat{t}=201$ for the compressed circuit and $\hat{t}=29$ for the best Trotter 
circuit. For the same system at timestep $t=2$, we find that at $\hat{\epsilon}=10^{-3}$ the 
compressed circuit has $\hat{t}=116$ while the best Trotter circuit has $\hat{t}=14$. At 
$\epsilon=10^{-2}$ we have $\hat{t}=378$ for the compressed circuit and $\hat{t}=46$ for 
the best Trotter circuit. From these values it is clear that for the chain we can go roughly 
eight times further in time than the best Trotter circuit with similar gate count. These 
values are for $L=12$, and the same analysis at $L=8$ reveals that here we can go fourteen 
to twenty times as far, while for $L=16$ we can go three to eight times as far, with the lower 
bounds for $t=2$ and the upper bounds for $t=1$. These values emphasize that the larger 
we choose $\hat{\epsilon}$, the larger the gap between $\hat{t}$ of the compressed and 
Trotter circuits becomes, which grows quadratically as stated above. This implies that 
the superiority of the compressed circuits over Trotter circuits becomes especially apparent 
when we set a relatively high error threshold, which for the compressed circuits is reached 
at much larger time than for Trotter circuits which have comparable gate count.

\begin{figure*}[t!]
	\centering
	\includegraphics[width=\textwidth]{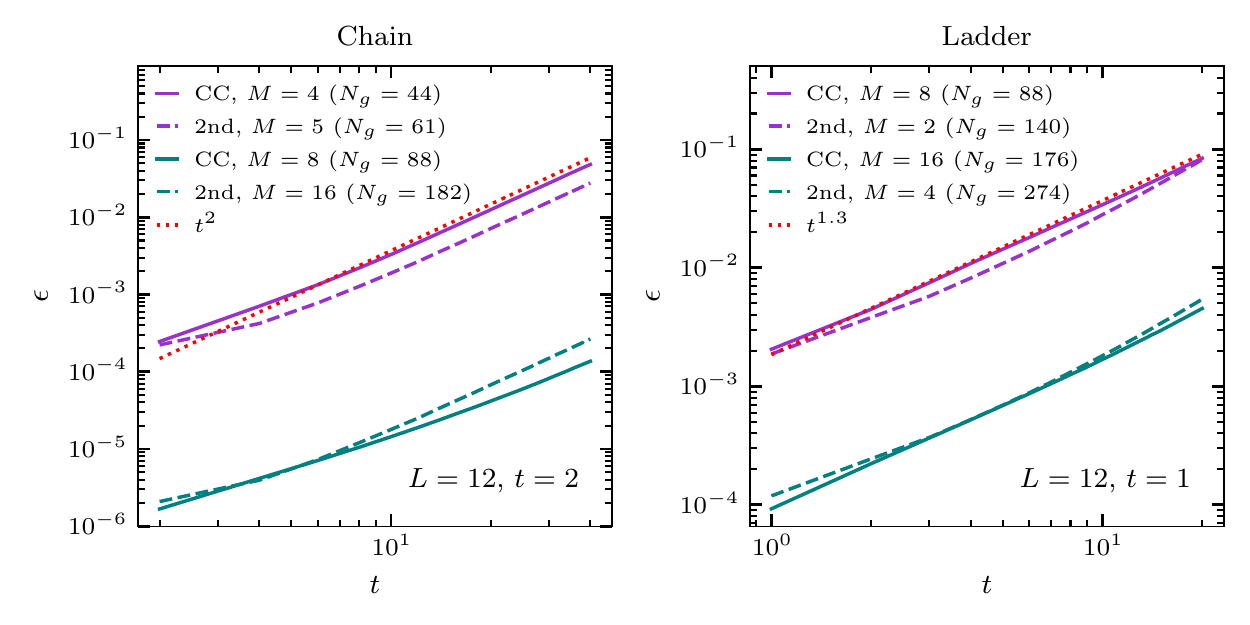}
	\caption{The infidelity $\epsilon$ versus stacking time $t$ for the time evolution 
	         operator of the $L=12$ Heisenberg model on a chain at $t=2$ (left panels) 
	         and ladder at $t=1$ (right panels), for compressed and second-order circuits 
	         that are stacked twenty times. The circuits were chosen such they have similar 
	         $\epsilon$ at the optimized $t$, with $M=4,8$ and $M=5,16$ for compressed and 
	         second-order Trotter circuits on the chain, and $M=8,16$ and $M=2,4$ for the 
	         ladder. As a result the compressed circuits have significantly lower gate count 
	         than the corresponding Trotter circuits. The red dashed lines are for the power 
	         laws $\epsilon\propto t^n$ with the best fitting power $n$.}
	\label{t_dependency_eqfidel}
\end{figure*}

Repeating this analysis for the ladder, again starting off with $L=12$ and $t=1$, we 
find at $\hat{\epsilon}=10^{-2}$ that the compressed circuit has $\hat{t}=34$ whereas 
the best Trotter circuit, excluding the fourth-order Trotter with double the gate count, 
has $\hat{t}=14$. With $\hat{\epsilon}=10^{-1}$ the compressed circuit has $\hat{t}=125$ 
whereas the second-order Trotter circuit has $\hat{t}=57$. For the same system at $t=2$ 
and with $\hat{\epsilon}=10^{-1}$, we have $\hat{t}=40$ for the compressed circuit and 
$\hat{t}=10$ for the second-order Trotter circuit. Hence for the ladder we can go roughly 
two to four times as far than the best Trotter circuit with comparable gate count. Repeating 
this analysis for $L=8$ we find that we can go five to two times farther, and for $L=16$ we 
can go three to two times farther, again with the lower bounds for $t=1$ and the upper bounds 
for $t=2$.

Instead of examining the stacking behavior of compressed and Trotter circuits with comparable 
gate count, we now compare how circuits with comparable optimized infidelity stack, to see 
whether similar fidelities are achievable with compressed circuits that have only a fraction 
of the gates of Trotter circuits. To this end we consider the chain and ladder for a single 
system size $L=12$, with time-step $t=2$ for the chain and $t=1$ for the ladder, and we stack 
the circuits up to $t=20$. For simplicity we compare only with second-order Trotter circuits, 
as we find analogous results for the other Trotter orders. For the chain we take compressed 
circuits with $M=4,8$, in which case the second-order Trotter circuits with similar 
optimized infidelity have $M=5,16$. Imporantly, while these compressed and Trotter circuits 
have similar fidelity, the $M=5$ Trotter circuit has $1.4$ times the gate count of the $M=4$ 
compressed circuit, whereas the $M=16$ Trotter circuit has $2.1$ times the gate count of the 
$M=8$ compressed circuit. For the ladder we take compressed circuits with $M=8,16$, such that 
the corresponding second-order Trotter circuits have $M=2,4$, i.e. they contain $1.6$ times as 
many gates. 

The results are displayed in Fig. \ref{t_dependency_eqfidel} in log-log scale, where in the 
left panel we show the stacked infidelities for the chain and in the right panel for the ladder. 
The red dashed lines are for the power laws $\epsilon\propto t^n$ with the best fitting power $n$. 
It is seen that the infidelity increases similarly for all considered pairs of compressed and 
Trotter circuits, which like Fig.~\ref{infidelity_stack_threshold} emphasizes that the 
compression strategy expounded in Sec. \ref{sec_method} has no drawbacks at long times, 
relative to the Trotter circuits. Moreover, the mentioned discrepancy in gate counts, with 
in all cases the Trotter circuit having significantly more gates, makes the compressed circuits 
especially favorable for simulation on real quantum devices, where the error due to gate 
imperfections and decoherence noise hampers time evolution.

\subsection{Out-of-time-ordered correlators}\label{subsec_otocs}

\begin{figure*}[h!]
    \centering
    \includegraphics[width=0.8\textwidth]{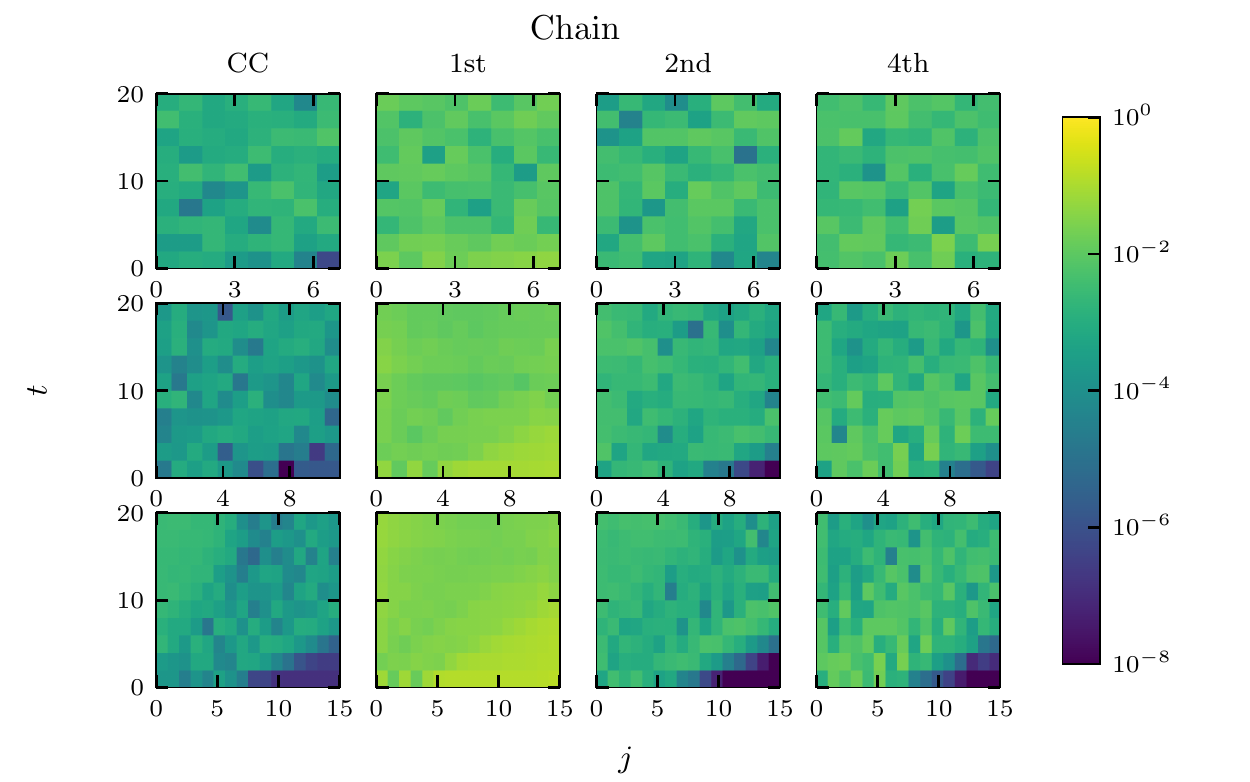}
    \includegraphics[width=0.8\textwidth]{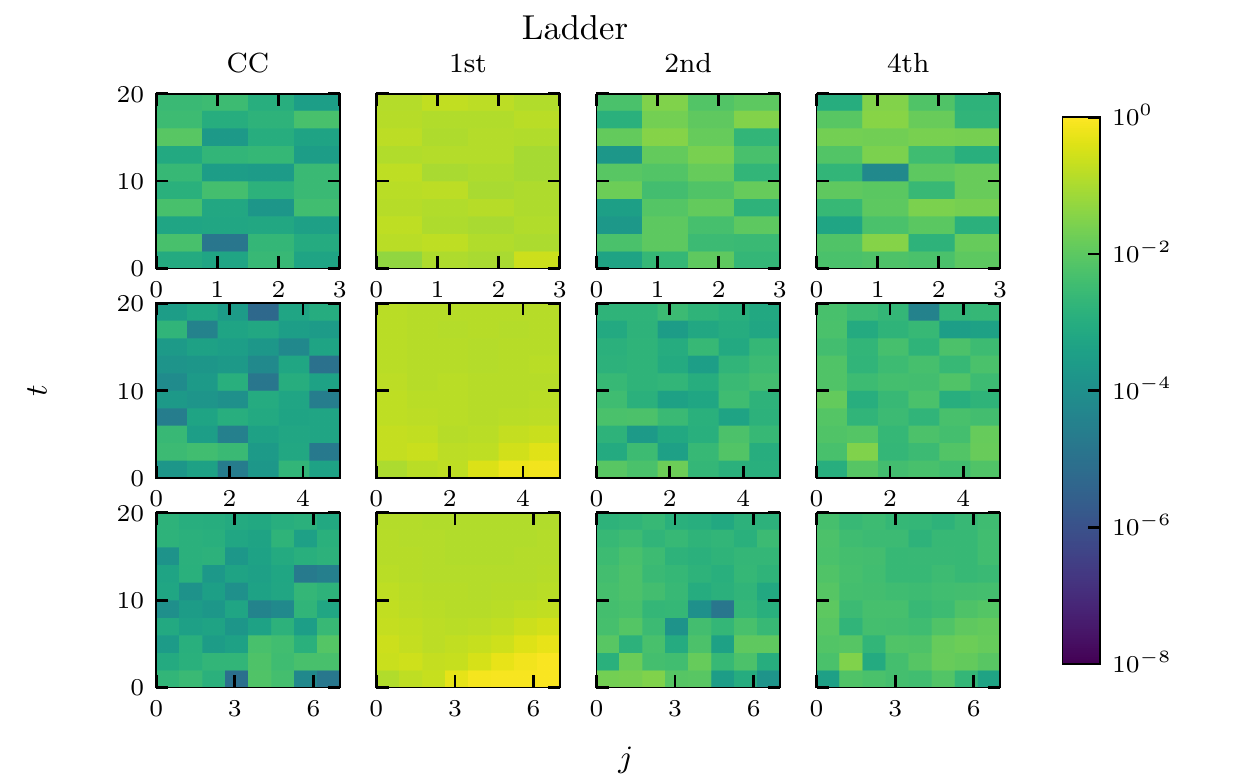}
    \caption{The absolute $C_{i=2,j}(t)$ errors for the chain (top three rows) and ladder 
             (bottom three rows) for a compressed circuit optimized at $t=2$ and stacked up 
             to ten times, along with the errors for Trotter circuits with similar gate counts. 
             For the chain $j$ labels the sites and for the ladder it labels the rungs. The first 
             and fourth row are for $L=8$ with $\chi=256$, the second and fifth row are for $L=12$ 
             with $\chi=150$, and the third and sixth row are for $L=16$ with $\chi=100$. The 
             first column is for the compressed circuit, the second, third and fourth columns are 
             for the first-, second- and fourth-order Trotter circuits. To have roughly equal gate 
             counts, the used depths are $M=8,8,7,1$ for the chain and $M=16,4,3,1$ for the ladder, 
             for the compressed circuit and first-, second- and fourth-order Trotter circuits, 
             respectively.}
    \label{mesh_OTOC_err_stack_dependency}
\end{figure*}

Having studied the infidelity and its behavior under stacking in detail in 
Sec. \ref{subsec_infidelity}, we now use the compressed circuits to determine the behavior of 
a quantity that does not enter the objective function (\ref{eq:fidelity_def}), namely the 
OTOC (\ref{otoc_definition}). 

In Fig. \ref{mesh_OTOC_err_stack_dependency} we show the absolute $C_{i=2,j}(t)$ errors, relative
to the targeted time-evolution operator, for compressed circuits which were optimized for $L=8,12,16$ 
chains and ladders at $t=2$ and stacked up to ten times, along with the errors for Trotter circuits 
with gate counts similar to these compressed circuits. For the chain we let $j$ run over all sites, 
whereas for the ladder it runs over all rungs. The upper three rows are for the chain while 
the lower three rows are for the ladder. The first and fourth row are for $L=8$, the second 
and fifth row are for $L=12$, and the third and sixth row are for $L=16$. The left column 
is for the compressed circuit while the second, third and fourth columns are for the first-, 
second- and fourth-order Trotter circuits. As in Fig. \ref{infidelity_stack_threshold} the 
depths are $M=8,8,7,1$ for the chain and $M=16,4,3,1$ for the ladder, for the compressed 
circuit and first-, second- and fourth-order Trotter circuits, respectively.

For the chain it is clear that the compressed circuit works better than the Trotter circuits 
within the lightcone, whereas it is slightly worse than the second- and fourth-order Trotter 
circuits at approximating the small values outside of the lightcone. For the ladder the 
compressed circuit is better everywhere, even better than the fourth-order Trotter circuit 
which has twice as many gates. Hence we draw the same conclusion as from 
Fig. \ref{infidelity_stack_threshold}: With a similar amount of gates we are able to go farther 
in time with the compressed circuits than with the Trotter circuits, before reaching some error 
threshold, even though we do not optimize based on OTOCs.

In Sec. \ref{otoc_errs} we show the OTOC values corresponding to the errors from Fig. 
\ref{mesh_OTOC_err_stack_dependency}, for compressed circuits and the targeted time-evolution 
operators. There we also show how the relative error of $C_{i=2,j=4}(t)$ propagates with stacking, 
for compressed and Trotter circuits that have similar optimized fidelity, indicating that we can 
maintain similar fidelity with compressed circuits that have a fraction of the amount of gates of 
the Trotter circuits.

\subsection{Analysis of the compressed circuit}\label{subsec_structure}

In the previous Sections \ref{subsec_infidelity} and \ref{subsec_otocs} we have seen that the 
compressed circuit outperforms the Trotter circuits. Here we investigate how this is achieved, 
by probing the structure of the layers and gates that make up the compressed and Trotter circuits.

\begin{figure}[h]
	\centering
    \includegraphics[width=\textwidth]{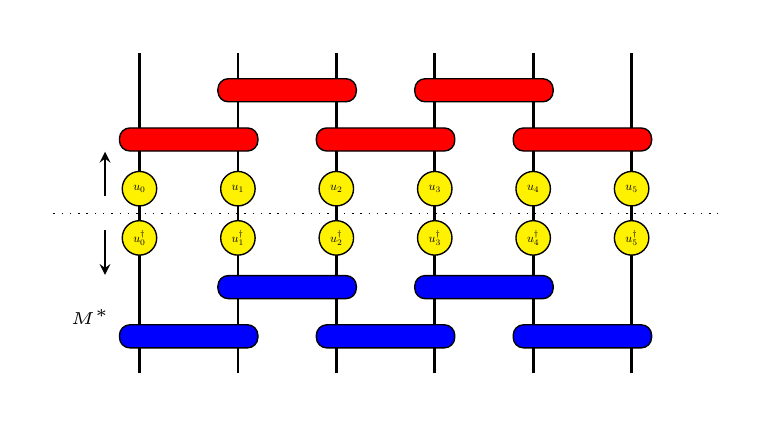}
    \caption{The gauge freedom that exists between the layers of a circuit. When we cut the 
    circuit across the horizontal dashed line, and want to use the lowest $M^*$ layers to 
    calculate an infidelity, we have to take into account the gauge freedom that is encoded 
    by inserting a pair of conjugate one-qubit unitaries $u_i^\dagger u_i=I$ at each qubit, 
    and absorbing one unitary upwards and the other downwards.}\label{fig:gauge_freedom}
\end{figure}

Starting off, we take a compressed circuit and Trotter circuits with comparable gate counts, 
and consider the infidelity between a subset of layers $M^*<M$ (counting from the bottom layer) 
and the time evolution operator at a time $t^*<t$ that is smaller than the time-step $t$ at 
which the compressed circuit was optimized. Crucially, we must take into account the gauge 
freedom that exists between layers, where we are able to insert conjugate layers of one-qubit 
unitaries, and absorb one layer into the subset we are considering and the other layer into its 
complement. This process is illustrated in Fig. \ref{fig:gauge_freedom}. Hence when calculating 
a subset infidelity for the compressed circuit, we add a layer of one-qubit unitaries between 
the subset and the time evolution operator at $t^*$, and minimize the infidelity with respect 
to these one-qubit unitaries. This way we account for the gauge freedom. 

\begin{figure*}[h]
	\centering
    \includegraphics[width=0.9\textwidth]{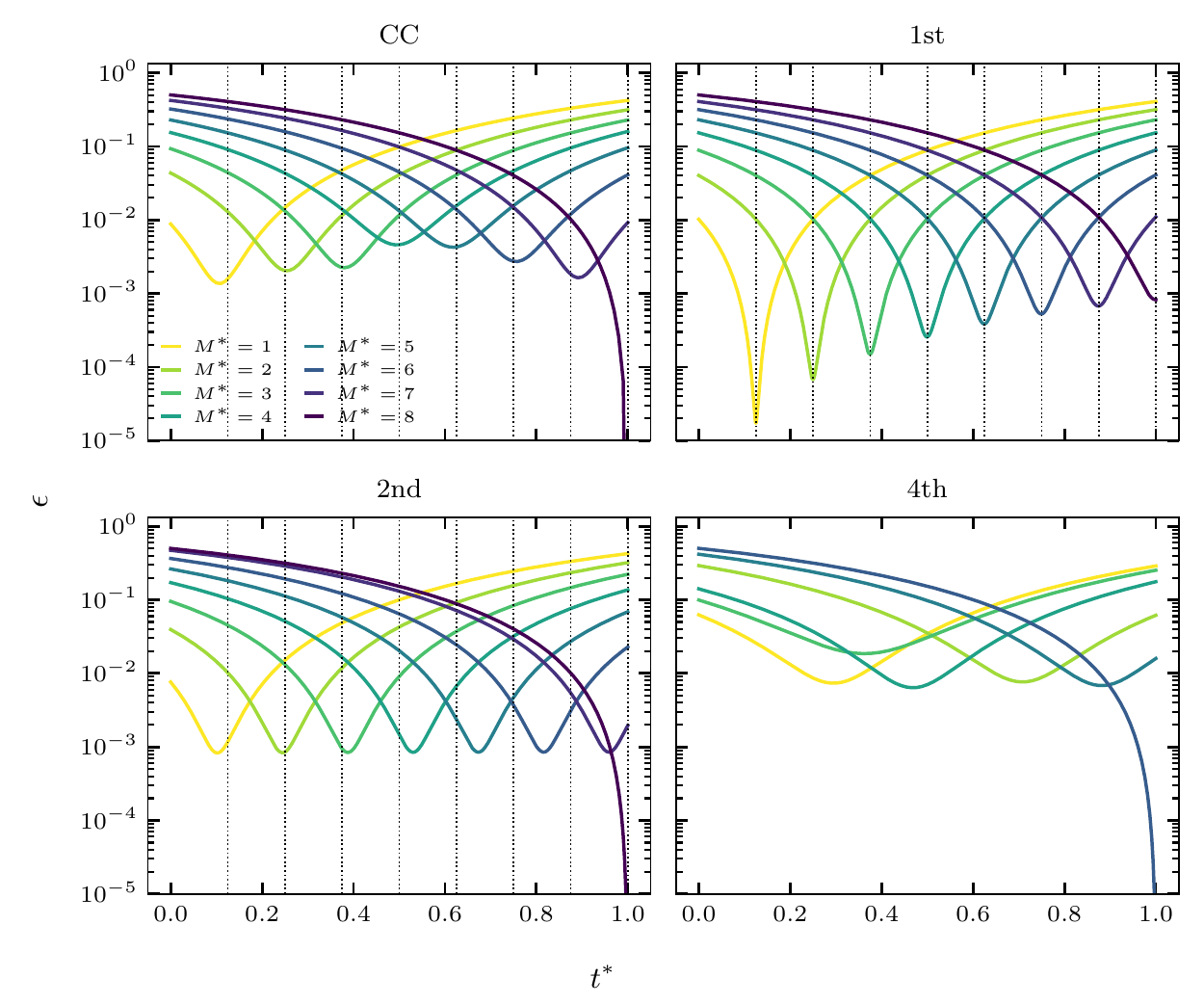}
	\caption{The infidelity between a subset of layers $M^*<M$, counting from the bottom layer, 
	         and the targeted time evolution operator at time $t^*<t$, where $t$ denotes the 
	         time-step at which the compressed circuit was optimized. The plots are for a 
	         Heisenberg chain with $L=8$ at $t=1$. In the top left panel we show the results 
	         for a compressed circuit with $M=8$, in the top right for a first-order Trotter 
	         circuit with $M=8$, in the bottom left for a second-order Trotter circuit with 
	         $M=7$, and in the bottom right for a fourth-order Trotter circuit with $M=1$. 
	         These depths were chosen such that the circuits have similar gate count. The curve 
	         with $M^*=M$ corresponds to the full circuit. The dashed lines mark times $tM^*/8$.}
	\label{layerwise_overlap_t1.0_NN}
\end{figure*}

In Fig. \ref{layerwise_overlap_t1.0_NN} we show the results for the chain with $L=8$ at $t=1$, 
for a compressed circuit with $M=8$ and Trotter circuits with $M=8,7,1$ for first-, second- 
and fourth-order, which have gate counts close to that of the compressed circuit. Here we 
define a Trotter circuit with $M^*$ layers as having $M^*$ brickwall layers, and the largest 
shown $M^*$ is the full circuit, which e.g. for the second-order Trotter circuit involves 
adding half a brickwall layer to its largest subset. For the compressed circuit $M^*=8$ 
corresponds to the full circuit. The dashed lines mark the times $t^*=tM^*/8$.

From Fig.~\ref{layerwise_overlap_t1.0_NN} it is clear that at $t=1$ there is significant overlap 
of the subsets with a time evolution operator at $t^*<t$ for both the compressed and Trotter 
circuits. However, in contrast to the first- and second-order Trotter circuits, where the 
infidelity dips are equidistant, and where for the first-order Trotter circuit the dip depth 
is decreasing with the number of stacked layers while for the second-order Trotter circuit it 
is constant, the dips of the compressed circuit are instead roughly symmetric and are smallest 
around $t^* \approx t/2$. A closer look reveals that the infidelity at this point is 
roughly $10^{-2}$, which is more than one order of magnitude larger than for the first- and 
second-order Trotter circuit at similar $t^*$. This is even more remarkable when taking the 
final infidelity into account, which is $\epsilon=1.8\cdot10^{-9}$ for the compressed circuit 
and therefore at least three orders of magnitudes better than the first-, second- and 
fourth-order Trotter circuits, which have $\epsilon=8.2\cdot10^{-4},1.2\cdot10^{-6},2.1\cdot10^{-6}$.
\begin{figure}[h]
	\centering
    \includegraphics[width=0.7\textwidth]{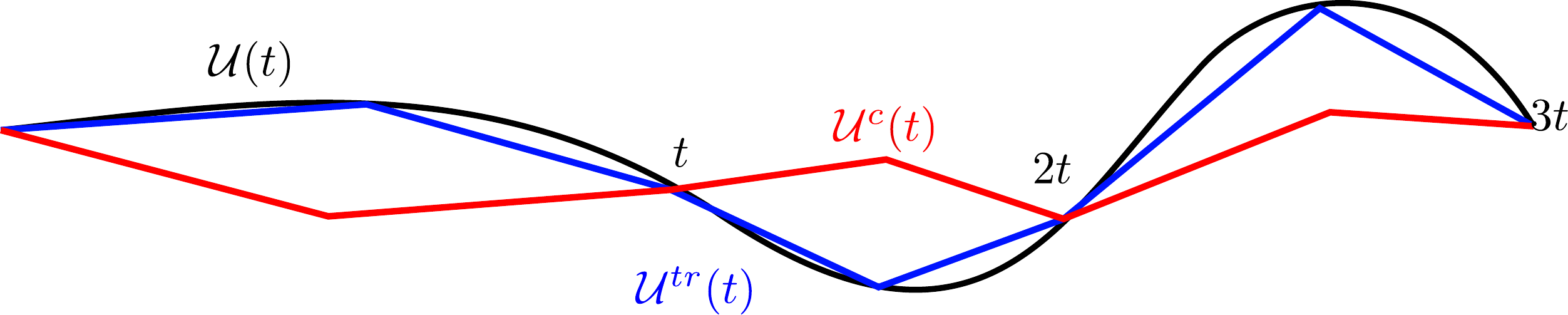}
	\caption{A sketch of the "refocussing" mechanism that potentially explains the structures 
	         observed in Fig. \ref{layerwise_overlap_t1.0_NN}. Here the targeted time evolution 
	         $\mathcal{U}(t)$ is shown in black, the Trotter evolution $\mathcal{U}^{tr}(t)$ is 
	         shown in blue, and the compressed evolution $\mathcal{U}^{c}(t)$ is shown in red. 
	         While $\mathcal{U}^{tr}(t)$ follows the target trajectory quite closely, 
	         $\mathcal{U}^{c}(t)$ instead becomes "refocussed" at multiples of the optimization 
	         timestep $t$.}\label{fig:refocus}
\end{figure}
This indicates that the compressed circuit does not follow the target "trajectory" given by the 
unitary time evolution, but slightly deviates from it. However, it becomes "refocused" 
at $t^*=t$, which we sketch in Fig. \ref{fig:refocus}. It is an interesting question for future 
research to understand the alternative trajectory, which might be beneficial for an optimal 
discretization of time evolution beyond the Trotter decomposition. In Sec. \ref{otoc_errs} we 
show that the refocussing also occurs for the OTOCs.

We note that we did not find these symmetric dips for all our compressed circuits, especially 
for larger $t$ and the ladder geometry. It remains an open question whether this is an artefact 
of the convergence of the optimization to a non-global minimum.

\begin{figure*}[h]
	\centering
    \includegraphics[width=0.9\textwidth]{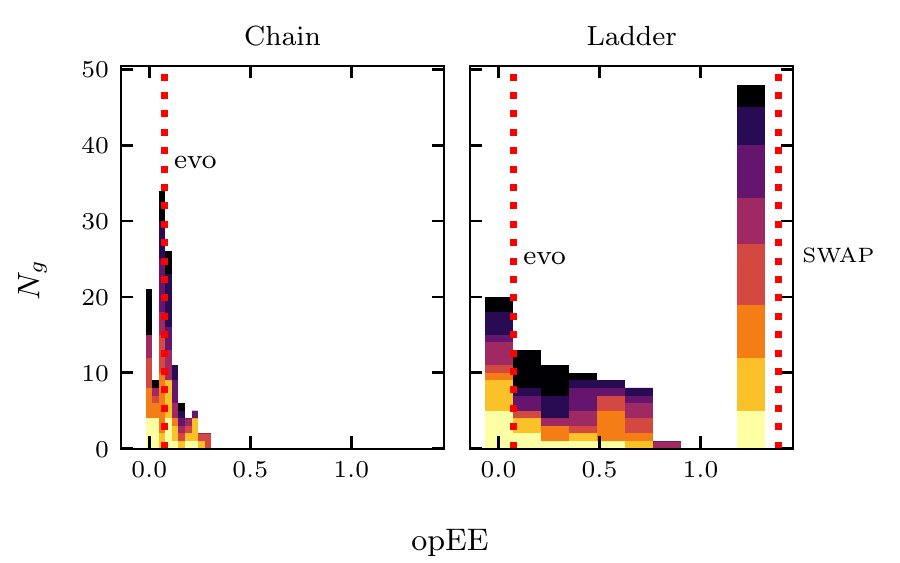}
	\caption{Stacked histograms for the opEE of the gates of a compressed circuit with 
	         depth $M=8$, optimized at $t=2$ for a $L=16$ chain (left panels) and ladder 
	         (right panels). The colors denote the contents of each layer, with the lightest 
	         color for the bottom layer and the darkest for the top layer. The red vertical 
	         lines denote the values for the gates in a $M=8$ first-order Trotter circuit, 
	         with the two lines in the ladder plots corresponding to the evolution and SWAP gates.}
	\label{gate_count_L16_M8_t2.0}
\end{figure*}

\begin{figure*}[h]
        \centering
        \includegraphics[width=.8\textwidth]{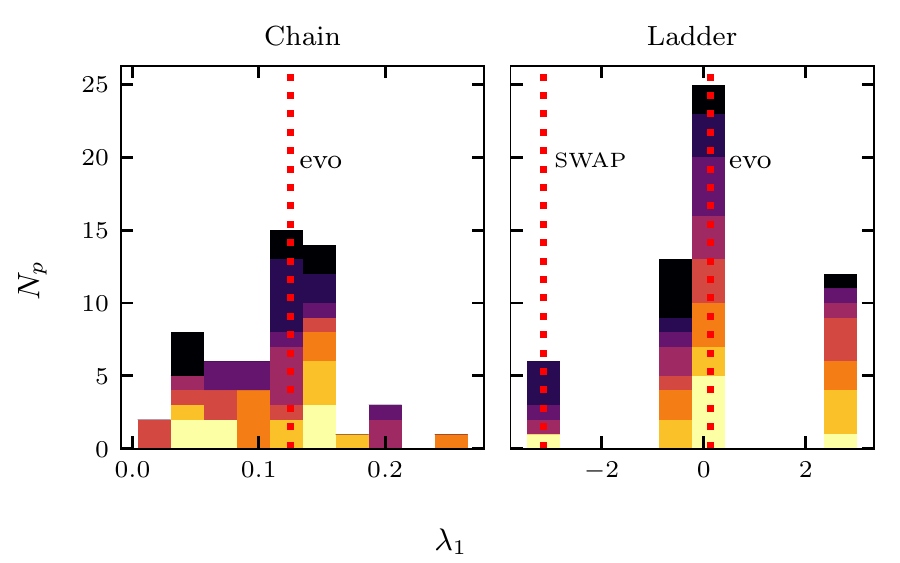}
        \caption{The distribution of the $\lambda_1$ parameter which enters the two-qubit unitary 
                 parameterization that was used in this work, shown for the chain (left panel) 
                 and ladder (right panel) with $L=8$ at $t=1$. The parameter count $N_p$ for a 
                 compressed circuit with $M=8$ is shown as a stacked histogram, with the lightest 
                 color corresponding to the bottom layer and the darkest color to the top layer. The 
                 first-order Trotter evolution gate value $\lambda_1^\mathrm{evo}=t/M$ and the SWAP 
                 gate value $\lambda_1^\mathrm{SWAP}=-\pi$ are shown as dashed red lines. The other 
                 two-qubit parameters $\lambda_2$ and $\lambda_3$ are distributed similarly. Note the 
                 different scales of the x-axes.}
        \label{lambda_histo_L8_M8_t1.0}
\end{figure*}

As a further comparison between compressed and Trotter circuits, we calculate the operator 
entanglement entropy (opEE) of their gates \cite{zhou_operator_2017, Prosen2007Operator }. 
Concretely, we take an optimized compressed circuit $\mathcal{C}$ and decompose each two-qubit 
gate $U_{ij}\in\mathcal{C}$ using a singular value decomposition into
\begin{equation}
U_{ij}=\sum_{l=1}^{4}s_lv^l_i\otimes v^l_j,
\end{equation}
where $v^l_i$ and $v^l_j$ are two sets of four one-qubit operators, acting on qubit $i$ and $j$ 
respectively, and where the four singular values $s_l$ encode the opEE of $U_{ij}$ as
\begin{equation}
\mathrm{opEE}=-\sum_l s_l^2\ln(s_l^2).
\end{equation}

In Fig. \ref{gate_count_L16_M8_t2.0} we display the opEE of all gates in a $M=8$ compressed 
circuit for the chain (left panel) and ladder (right panel) for $L=16$ at $t=2$. The histograms 
are stacked, with each color denoting the content of a layer, where the lightest color represents 
the bottom layer and the darkest color the top layer. The red vertical lines mark the values for 
the $M=8$ first-order Trotter circuit, with the two lines in the ladder plots corresponding to 
the evolution and SWAP gates. These histograms show that the gates of the compressed circuit are 
more hetergenous compared to those of the Trotter circuits, since they have a relatively large 
spread in opEE instead of one or two values. Moreover, for the ladder it is seen that a several 
gates in the compressed circuit assume an opEE that is near to that of the SWAP gate, which we 
view as an indication that the action of the SWAP gate is baked into our optimized circuits.

Finally we consider the distribution of the parameter $\lambda_1$ across the optimized two-qubit 
unitaries, which are parameterized as in (\ref{twoqubit_param}). We found that $\lambda_2$ and 
$\lambda_3$ are distributed similarly. In Fig. \ref{lambda_histo_L8_M8_t1.0} we show histograms 
for the parameter counts $N_p$ of $\lambda_1$ for the chain (left panel) and ladder (right panel) 
with $L=8$ at $t=1$, for a compressed circuit with $M=8$. Note here the different scales of the 
x-axes. The histograms are again stacked, with the lightest color corresponding to the bottom 
layer and the darkest color to the top layer. The red dashed lines mark the values of the gates 
in the $M=8$ first-order Trotter circuit, for which $\lambda_1^\textrm{SWAP}=-\pi$ and 
$\lambda_1^\textrm{evo}=t/M$, both having no one-qubit dressing (\ref{onequbit_param}). As in 
Fig. \ref{gate_count_L16_M8_t2.0}, we see that the gates of the compressed circuit have a larger 
spread than the gates of the Trotter circuit, which instead assume one or two values. Also, for 
the ladder we again observe an accumulation of gates near the SWAP value.

The gates appearing in the optimized circuits appear to encode more structure than gates from 
Trotter circuits and are generally speaking encoding a larger change of the wave function per 
gate compared to the case of Trotter circuits. This can be seen best in the limit of very small 
Trotter time steps, in which each appearing gate (except SWAP) is very close to identity, while 
in the opposite limit which we optimize for, each gate needs to be sufficiently different from 
identity in order to represent the same time evolution operator.

\section{Conclusion and Outlook}\label{sec_conclusion}

In this work we have presented an approach which reduces the resource cost of digital quantum 
simulation compared to standard Trotter decompositions by globally optimizing a simple 
parameterized brickwall circuit in a way that is scalable to large systems. Crucially, 
the performance per gate is better even when the compressed circuit does not respect the 
connectivity of the simulated lattice, potentially allowing for high fidelity simulation of 
systems with a connectivity that is larger than that of the used quantum processor. To illustrate 
this we have compared the infidelity of the compressed and Trotter circuits with the targeted time 
evolution operators of Heisenberg chains and ladders, as well as the ability to reproduce their 
OTOCs.

We have shown that we can achieve similar accuracy of the time evolution operator with up to one 
order of magnitude less gates, depending on the desired accuracy and system. Moreover, we checked 
that this advantage persists when stacking the circuits many times, a central ingredient to 
simulating a quantum system over long times. This enables high fidelity propagation to times 
which are currently elusive with conventional Trotter decomposition methods. 

Furthermore, we analyzed the structure of the compressed circuits. In the case of the chain, 
we observed a "refocussing" mechanism, which suppresses the infidelity at multiples of the 
optimized time step, while the evolution inside the optimized circuit appears to follow a 
trajectory which is further away from the targeted time evolution operator. It is an interesting 
question for further research to understand this trajectory and relate it also to recent studies 
of Trotter decompositions and its breakdown for large time steps~\cite{Heyl2019Quantum,Kargi2021Quantum}.

Our results open the door for many further directions. As a next step, one can for example take 
symmetries into account to further reduce the number of parameters. This might be especially 
favorable when exploiting translation symmetries. Furthermore, one can optimize the circuits 
with other cost functions than the fidelity, as was also done for example 
in~\cite{Bolens2021Reinforcement}. Promising directions are using local observables or 
density matrices. While such an approach might simplify the convergence of the optimization, 
it is still an open question to what extent the accurate simulation of observables or other 
general quantities would be recovered.  

We end by stressing that in this work we have used the simplest possible noise model, 
by assuming that each applied gate introduces the same amount of noise to the system and that 
therefore a minimization of the gate count reduces the overall noise. A refinement of this 
noise model will be the subject of future research.

\section*{Acknowledgments}
We thank Luis Colmenarez for useful comments on the manuscript. D.H. thanks Adam Smith, 
Frank Pollmann, and Hongzheng Zhao for useful discussions.
\paragraph{Funding information}

This project was supported by the Deutsche Forschungsgemeinschaft (DFG) through SFB 1143 
(project-id 247310070) and the cluster of excellence ML4Q (EXC 2004, project-id 390534769). 
We also acknowledge support from the QuantERA II Programme that has received funding from 
the European Union’s Horizon 2020 research innovation programme (GA 101017733), and
from the Deutsche Forschungsgemeinschaft through the project DQUANT (project-id 499347025).

\begin{appendix}

\section{Gate count equations}\label{sec_gate_counts}

Here we state the equations for the NN two-qubit gate counts $N_g$ of the first-, second- and 
fourth-order Trotter circuits of depth $M$, which are used in Sec. \ref{sec_results}. These 
are denoted by $N^{\mathrm{1st}}_{g\mathfrak{c}/\mathfrak{l}}(M)$, $N^{\mathrm{2nd}}_{g\mathfrak{c}/\mathfrak{l}}(M)$ and $N^{\mathrm{4th}}_{g\mathfrak{c}/\mathfrak{l}}(M)$, 
respectively, where $\mathfrak{c}$ corresponds to the chain and $\mathfrak{l}$ to the 
triangular ladder. In deriving these equations we made maximal use of the ability to combine 
gates in subsequent Trotter steps. The compressed circuits have gate count 
$N^{\mathrm{1st}}_{g\mathfrak{c}}$.

For the chain the equations are 
\begin{align}
N^{\mathrm{1st}}_{g\mathfrak{c}}(M) &= M(L-1), \\
N^{\mathrm{2nd}}_{g\mathfrak{c}}(M) &= M(L-1) + \floor{\frac{L}{2}}, \\
N^{\mathrm{4th}}_{g\mathfrak{c}}(M) &= 5M(L-1) + \floor{\frac{L}{2}}.
\end{align}
For the ladder, in which case we have to take into account the SWAP gates, the corresponding 
equations are
\begin{align}
    N^{\mathrm{1st}}_{g\mathfrak{l}}(M) &= M\left(4L - 7\right) \\
    N^{\mathrm{2nd}}_{g\mathfrak{l}}(M) &= 2M\left(N^{\mathrm{1st}}_{g\mathfrak{l}}(1) + 1\right) - (3M - 1)\floor{\frac{L}{2}} \\
    N^{\mathrm{4th}}_{g\mathfrak{l}}(M) &= 5M N^{\mathrm{2nd}}_{g\mathfrak{l}}(1) - (5M - 1)\floor{\frac{L}{2}}.
\end{align}

\section{Convergence of the optimization}\label{sec:convergence}

In order to find the optimal compressed circuit using the gradient descent method outlined in 
Sec. \ref{sec:optimization}, it is important to scan the hyperparameter space of the used optimizer. 
The reason is that there is no single set of hyperparameters which finds the best solution for all 
optimization problems. We find the best convergence by using the vanilla Adam optimizer \cite{adam_kingma_2014}, 
which is presented in Algorithm \ref{adam_algo}.

\begin{algorithm}
	\caption{\textit{Vanilla Adam \cite{adam_kingma_2014}.} This gradient-descent optimizer updates 
	         the circuit parameters $\vec{\theta}$ to minimize the infidelity $\epsilon(\vec{\theta})$, 
	         by taking into account exponentially decaying running averages of the first moment $m$ and 
	         second moment $v$ of the infidelity gradient $g$ for each parameter separately. Instead of 
	         choosing the parameter updates to be proportional to $g$, as in vanilla gradient descent, 
	         here it is proportional to a memory of the previous gradients $m$. This results in a 
	         relatively stable minimization and to some extent prevents getting stuck in local 
	         minima. Moreover, since the optimization algorithm is first order, the magnitude of the 
	         parameter update is proportional to its uncertainty in decreasing the infidelity. For this 
	         reason, large updates are undesirable, whereas tiny updates are also undesirable since they 
	         halt the minimization and promote getting stuck in local minima. With this in mind, the update 
	         magnitude is forced to be desirable, by choosing it to be proportional to $m/\sqrt{v}$.}\label{adam_algo}
\begin{algorithmic}
	\State \textbf{Hyperparameters:}
	    \State \quad $\lambda$: Raw learning-rate
	    \State \quad $\beta_1$: First moment decay strength
	    \State \quad $\beta_2$: Second moment decay strength
	    \State \quad $\delta$: Regularization
	    \State \quad $N_\text{iters}$: Amount of iterations
	\State \textbf{Initial conditions:}
      \State \quad $m_0 \gets 0$ (First moment initially zero)
      \State \quad $v_0 \gets 0$ (Second moment initially zero)
\For {( $i=0$; $i < N_\text{iters}$; i = i+1 )}
    \State $g_i \gets \nabla_{\vec{\theta}_{i-1}} \epsilon(\vec{\theta}_{i-1})$ (Calculate gradient at current parameters)
    \State $m_i \gets \beta_1 m_{i-1} + (1-\beta_1)g_i$ (Extend running average of first moment)
    \State $m^*_i \gets m_i/(1-\beta_1^i)$ (Bias correction)
    \State $v_i \gets \beta_2 v_{i-1} + (1-\beta_2)g_i^2$ (Extend running average of second moment)
   	\State $v^*_i \gets v_i/(1-\beta_2^i)$ (Bias correction)
    \State $\vec{\theta}_i \gets \vec{\theta}_{i-1}-\lambda m^*_i/(\sqrt{v^*_i} + \delta)$ (Update parameters)
\EndFor
\State \Return $\vec{\theta}_i$ (Final circuit parameters)
\end{algorithmic}
\end{algorithm}

\begin{figure*}[h!]
	\centering
    \includegraphics[width=0.7\textwidth]{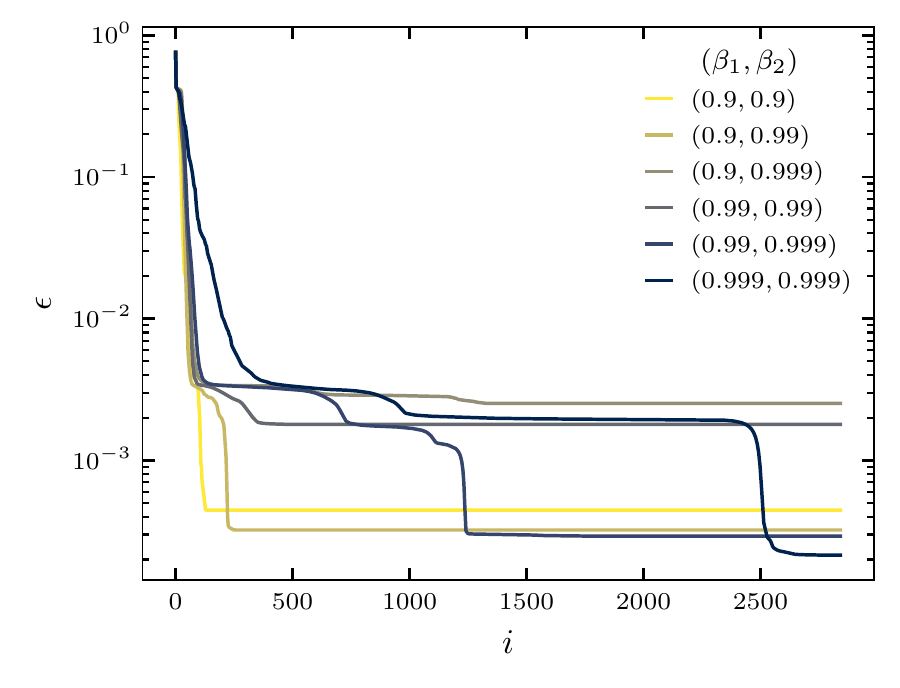}
	\caption{The infidelity $\epsilon$ as a function of the iteration step $i$ for an 
	         Adam optimizer with learning-rate $\tau=0.01$, regularization $\delta=10^{-4}$, 
	         and various decay rates $(\beta_1, \beta_2)$ with $\beta_1,\beta_2\in\{0.9, 0.99, 0.999\}$. 
	         The optimization is performed for a size $L=8$ ladder at time $t=1$ with circuit depth 
	         $M=8$. The lowest infidelity is reached with $(0.999,0.999)$, but crucially this requires
	         the optimizer to spend time in local minima without getting stopped by a convergence criterion 
	         when the infidelity has stagnated.}
	\label{fig:adam_convergence}
\end{figure*}

We scan the hyperparameter space $(\lambda, \delta, \beta_1, \beta_2)$ for the most favorable convergence properties. 
As mentioned in Sec. \ref{sec:optimization}, it is crucial to continue iterating the algorithm when we reach a 
plateau in the fidelity. This is illustrated in Fig. \ref{fig:adam_convergence}, where we display the gradient 
descent of $\epsilon$ for a circuit with $M=8$ layers on the time evolution operator of an $L=8$ ladder at $t=1$, and 
consider various $(\beta_1,\beta_2)$ with learning-rate $\tau=0.01$ and regularization $\delta=10^{-4}$. 
Here the largest fidelity is obtained with $\beta_1=\beta_2=0.999$, but we have to overcome multiple plateaus, 
which would have been spoiled by using a convergence criterion.

\section{OTOC details} \label{otoc_errs}

\begin{figure*}[t!]
	\centering
	\includegraphics[width=0.9\textwidth]{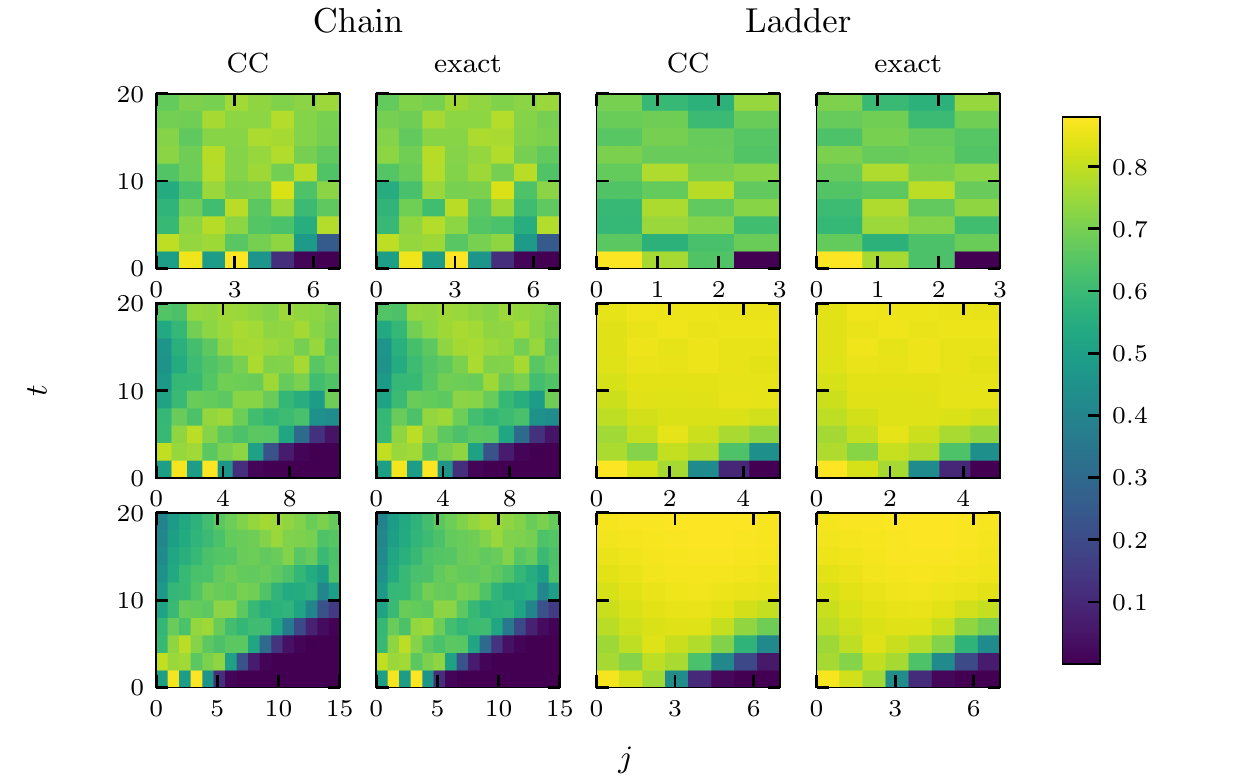}
	\caption{The OTOCs $C_{i=2,j}(t)$ as a function of site or rung $j$ and stacking time $t$
           for the chain (left columns) and the ladder (right columns), for compressed circuits 
           optimized at $t=2$ and stacked up to ten times (first and third columns) and the 
           corresponding target values (second and fourth columns). For the chain we take $M=8$ 
           and for the ladder $M=16$. The top row is for $L=8$ with $\chi=256$, the middle row 
           is for $L=12$ with $\chi=150$, and the bottom row is for $L=16$ with $\chi=100$.}
	\label{mesh_OTOC_stack_dependency}
\end{figure*}

\begin{figure*}[h]
	\centering
    \includegraphics[width=0.9\textwidth]{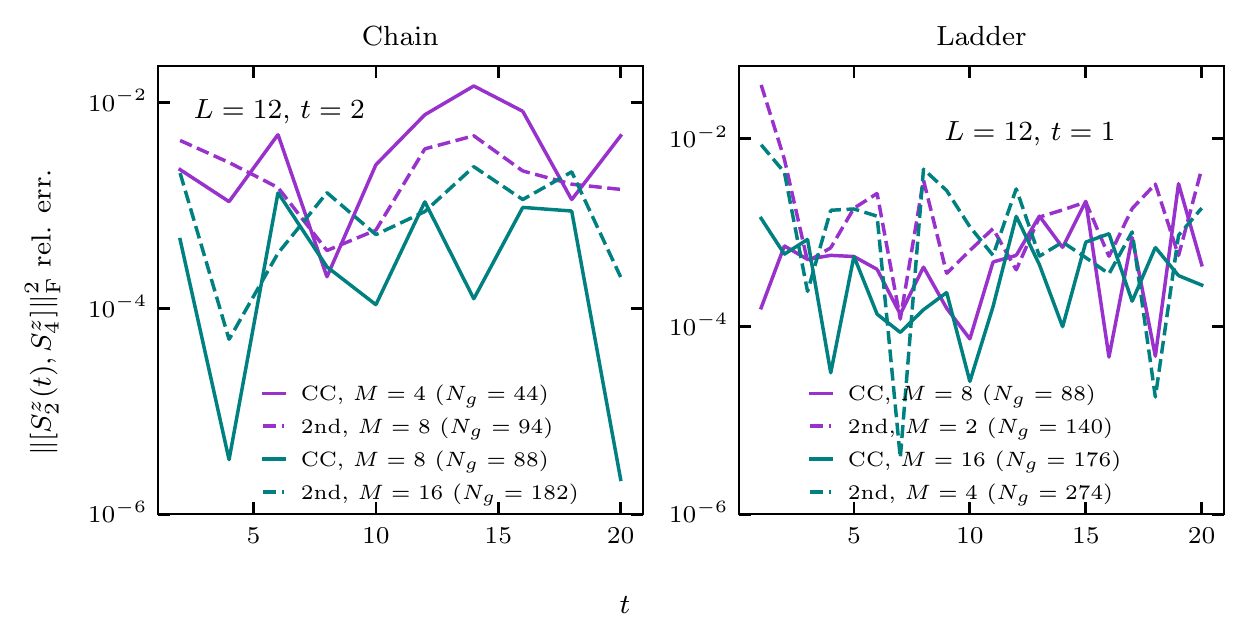}
	\caption{The relative error of the OTOC $C_{i=2,j=4}(t)$ versus stacking time $t$ for the chain (left panel) and ladder (right panel) with $L=12$, for circuits optimized at $t=2$ for the chain and $t=1$ for the ladder, and stacked up to time $t=20$. For the chain we consider $M=4,8$ and for the ladder $M=8,16$. For each $M$ we choose a second-order Trotter circuit with similar fidelity at the optimized $t$, i.e. $M=5,16$ for the chain and $M=2,4$ for the ladder. As a result the compressed circuits have significantly less gates than the corresponding Trotter circuits.}
	\label{OTOC_stack_dependency}
\end{figure*}

\begin{figure*}[h]
	\centering
    \includegraphics[width=0.7\textwidth]{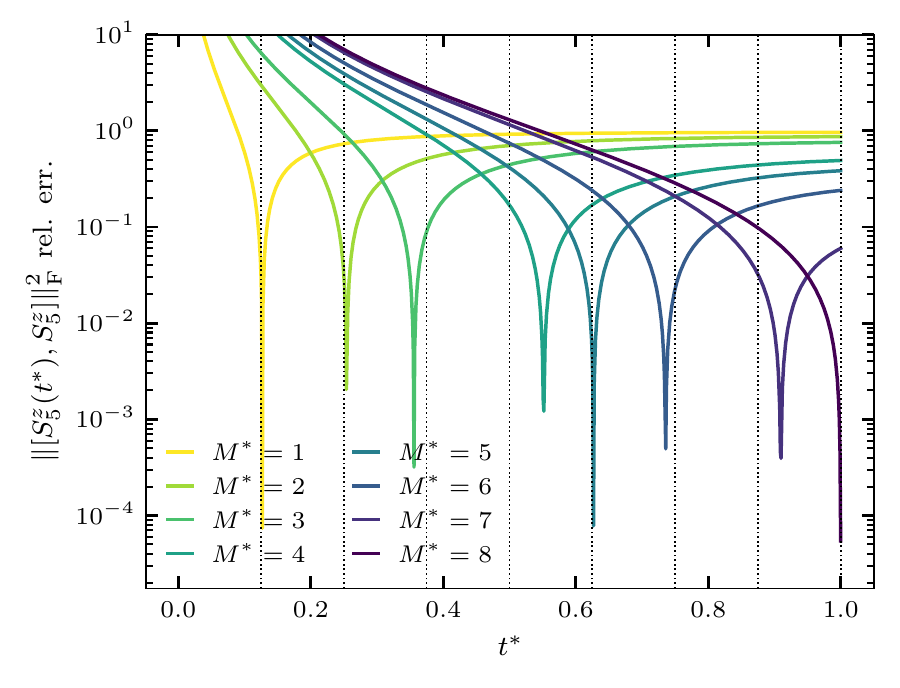}
	\caption{The relative error of the OTOC $C_{i=5, j=5}(t^*)$ between that of a subset $M^*$ 
	         of layers and that of the target unitary at time $t^*$, for the Heisenberg chain 
	         at $L=8$ and a depth $M=8$ circuit optimized at time $t=1$. The curve with 
	         $M^*=M$ corresponds to the full circuit and the dashed lines mark times $tM^*/8$.}
	\label{otoc_refocussing}
\end{figure*}

First we display the OTOC values of the stacked compressed circuits and targeted time-evolution 
operators that were used to make Fig. \ref{mesh_OTOC_err_stack_dependency}. In the left two panel 
columns of Fig. \ref{mesh_OTOC_stack_dependency} we show the OTOCs $C_{i=2,j}(t)$ for the chain 
and in the right two columns for the ladder. The first and third columns are for the compressed 
circuits, whereas the second and fourth columns are for the target unitaries. As already became 
apparent from Fig. \ref{mesh_OTOC_err_stack_dependency}, the agreement is excellent for all 
considered stacking times $t$. 

Now we consider the analog of Fig. \ref{t_dependency_eqfidel} for the relative error of the OTOC 
$C_{i=2,j=4}$. In particular, we consider the chain and ladder with $L=12$ and take a couple 
compressed circuits for which the infidelities were optimized at $t=2$ for the chain and $t=1$ 
for the ladder, which we then stack up to $t=20$. As in Fig. \ref{t_dependency_eqfidel} we take 
compressed circuits with $M=4,8$ for the chain and $M=8,16$ for the ladder, and we compare these 
with second-order Trotter circuits that have similar fidelity at the optimized time step, 
corresponding to $M=5,16$ for the chain and $M=2,4$ for the ladder. In Fig. 
\ref{OTOC_stack_dependency} we show the results, with the left panel for the chain and the right 
panel for the ladder. The implications are the same as those derived from Fig. 
\ref{t_dependency_eqfidel}: With a smaller amount of gates we essentially get the same performance, 
in this case even for a quantity that does not appear in the objective function (\ref{eq:fidelity_def}).

Finally, we check whether the refocussing that was observed for the infidelity $\epsilon$ in Fig. 
\ref{layerwise_overlap_t1.0_NN} also emerges for the OTOCs, which contrary to $\epsilon$ does 
not enter the cost function of the optimization scheme. In Fig. \ref{otoc_refocussing} we show 
the relative error of $C_{i=5,j=5}(t^*)$ between that of $M^*$ layers and that of the target unitary 
at time $t^*$. Before using the $M^*$ layers to calculate the OTOC at $t^*$, we minimize its 
infidelity with respect to the target unitary at $t^*$, taking into account the gauge invariance.
As in Fig. \ref{layerwise_overlap_t1.0_NN}, we perform the calculations for the Heisenberg chain 
with $L=8$ and a $M=8$ circuit optimized at $t=1$, with the results shown in Fig. 
\ref{otoc_refocussing}. We see that a similar refocussing takes place, with the minima 
for $M^*<M$ being elevated with respect to that at $M^*=M$ and with unequal spacing in time.

\section{Emergence of lattice symmetries}\label{sec:lattice_symmetries}

The brickwall circuit ansatz (\ref{eq:circ2}) used in this work has the most general form, 
consisting of arbitrary two-body unitaries and not taking into account any symmetry of the 
targeted time-evolution operator, i.e. in our case those corresponding to the Hamiltonians 
(\ref{hamiltonians}). To restrict the ansatz space it could be useful to incorporate such 
symmetries into the circuit at the gate level. 

Take for example the Heisenberg chain in (\ref{hamiltonians}),
which possesses lattice inversion symmetry, being invariant under a flip of the lattice across 
the middle bond for even $L$. To incorporate this into the ansatz we let the gate acting on the 
bond between sites $i$ and $i+1$ also act on the mirrored bond between $L-2-i$ and $L-1-i$, albeit
flipped across the time axis. Since this gate and its flipped counterpart should be equal for the 
inversion symmetry to be manifest, the gate parameterization (\ref{gate_param}) implies that 
the one-qubit unitary $u_i$ should be equal to $u_j$, and that $u'_i$ should be equal to $u'_j$, 
with the two-qubit unitary $v_{ij}$ being flip-symmetric by construction.

\begin{figure*}[h!]
	\centering
    \includegraphics[width=\textwidth]{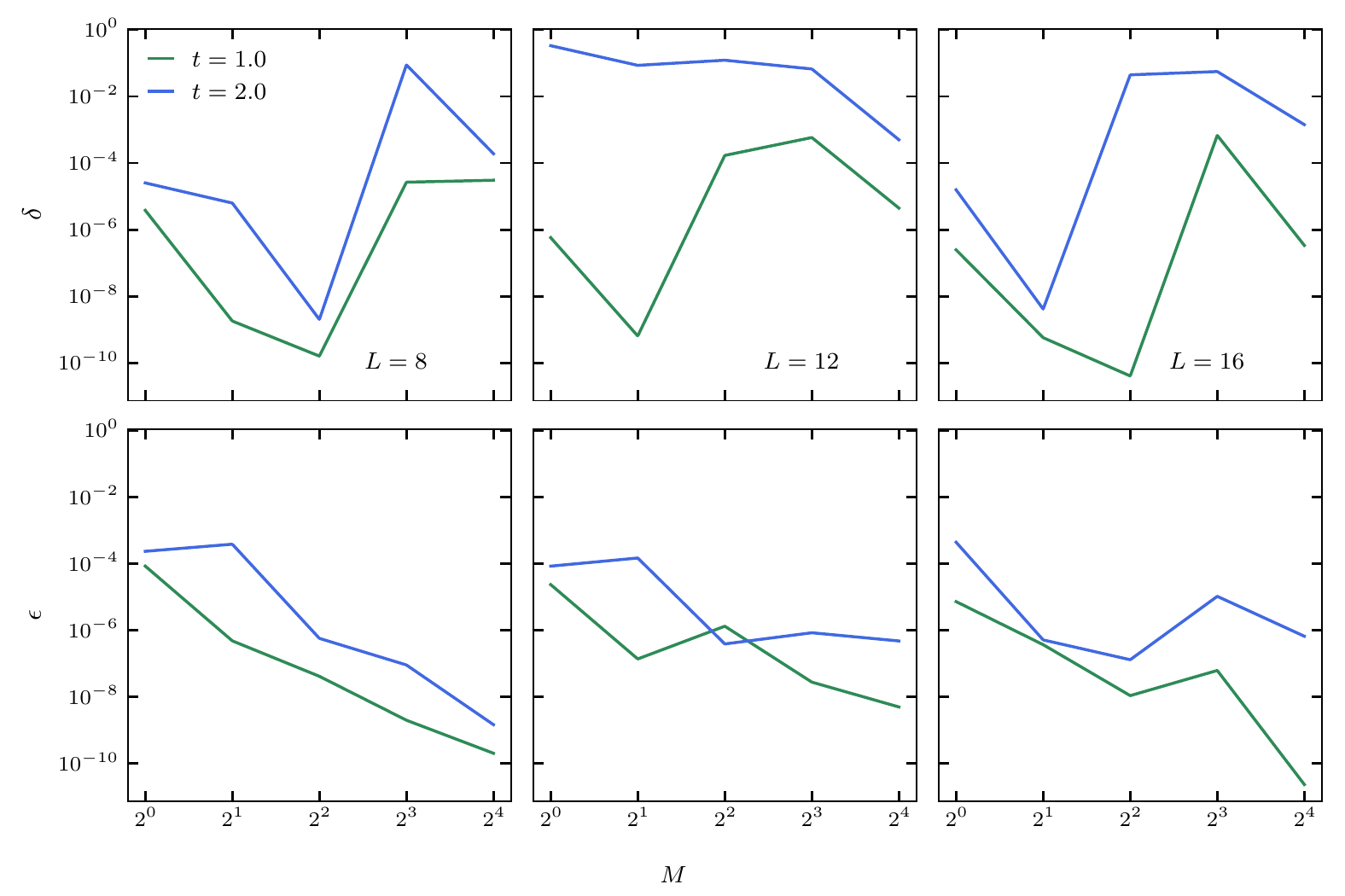}
	\caption{The average gate-wise infidelity $\delta$ of every gate with its mirrored counterpart (top panels), 
	         flipped across the middle bond, for all gates in compressed circuits which were 
	         optimized to approximate the lattice inversion symmetric Heisenberg chain time-evolution 
	         operator. For comparison, we also show the infidelity $\epsilon$ of the 
	         circuit as a whole with its mirrored counterpart (bottom panels). These quantities probe
	         to which extent the inversion symmetry of the targeted unitary emerges in the compressed 
	         circuit. The infidelities are shown as a function of the circuit depth $M$, for times 
	         $t=1$ and $t=2$. The left panels are for system size $L=8$, the middle panels are for 
	         $L=12$, and the right panels are for $L=16$. }
	\label{invsym_infidelity}
\end{figure*}

Since we did not incorporate this inversion symmetry into the circuits used for our simulations, 
it is an interesting question whether the chosen circuit ansatz in combination with the 
optimization procedure leads to its emergence. To probe this, we take an optimized circuit and 
for each of its gates we calculate the infidelity with its mirrored counterpart, and then average 
over all gates to get the average gate infidelity $\delta$. As for the subset infidelity 
from Fig. \ref{layerwise_overlap_t1.0_NN}, here it is crucial to take into account the gauge 
symmetry. We also calculate the infidelity $\epsilon$ of the circuit as a whole with its mirrored counterpart, 
to determine if it is reasonable to expect the symmetry to emerge on the gate level. If this 
overall infidelity is high, it is unlikely that it is low at the gate level. The results are 
shown in Fig. \ref{invsym_infidelity}.

\end{appendix}

\clearpage

\bibliography{references}

\nolinenumbers

\end{document}